\documentclass[reprint,amsmath,amssymb,aps]{revtex4-1}
\usepackage{graphicx}
\usepackage{dcolumn}
\usepackage{bm}
\usepackage[dvipsnames]{xcolor}

\definecolor{amethyst}{rgb}{0.54, 0.17, 0.89}
\definecolor{coral}{rgb}{1.0, 0.3, 0.4}

\usepackage{amssymb}
\usepackage{amsmath}
\usepackage{epsfig}
\usepackage{chemarr}
\usepackage{url}
\usepackage{natbib}
\usepackage{color}

\usepackage[defaultcolor=red, final]{changes}

\begin{document}
\title{Contrastive  learning in tunable dynamical systems}

\author{Menachem Stern$^1$, Adam G. Frim$^2$, Raúl Candás$^1$,  Andrea J. Liu$^{2,3}$ and Vijay Balasubramanian$^{2,3}$}

\affiliation{$^1$AMOLF, Science Park 104, 1098 XG Amsterdam, The Netherlands}

\affiliation{$^2$Department of Physics and Astronomy and Center for Soft and Living Matter, University of Pennsylvania, Philadelphia, PA 19104}

\affiliation{$^3$Santa Fe Institute, 1399 Hyde Park Road,
Santa Fe, NM 87501, USA}

\date{\today}

\begin{abstract} 
\noindent We generalize the theory of supervised contrastive learning, previously applied to physical systems at equilibrium or steady state, to systems following any dynamics described by coupled ordinary differential equations.  We show that if physical dynamics break time reversal symmetry, gradient descent on a cost function embodying the desired behavior cannot be achieved with a scalable process, even in principle.  We therefore introduce Probably Approximately Right (PAR) learning processes, composed of a local contrastive learning rule and a scalable supervision protocol. We show that approximate, local supervision with forward propagation of the error signal can be used to successfully train  several tunable models of physical dynamics inspired by examples in biological and machine learning. 
\end{abstract}

\maketitle

\section{Introduction}

Living systems adapt themselves over timescales ranging from seconds  to years~\cite{folse2010individual}, in order to manage changing circumstances or to develop new collective behaviors. Indeed, the various organisms themselves are the results of genome tuning by biological evolution on very long time scales. Within an organism, different systems can be adapted to the niche of the species, environmental conditions, interoceptive state, and functional requirements -- consider, for example, the different organizations and dynamics of the adaptive immune systems of vertebrates versus bacteria \cite{mayer2015well,mayer2015well,mayer2016diversity,bradde2020size}.
Evolvability and functional adaptation to external signals~\cite{stern2023learning} both underpin the resilience of life and individuality~\cite{kandel1992biological,sterling2015principles}, and can be considered together within the broad context of tunable physical systems. Here, by saying that a system is ``tunable" we mean that it has many degrees of freedom characterizing interactions that can be individually adjusted to give rise to complex collective behaviors.

In some cases living systems rely on local comparisons between different functional states to guide the tuning of such degrees of freedom. For example, spike-timing-dependent plasticity (STDP) in neural synapses operates on the relative timing of action potentials in pre- and post-synaptic neurons~\cite{caporale2008spike,bengio2015stdp, richards2019deep}. In the slime mold {\it Physarum Polycephalum}, vascular diameters are determined by comparison of wall shear stress between an excited (time delayed) and a time-averaged state~\cite{whiting2016towards,le2023physarum,marbach2023vein}. Likewise, in the actin cytoskeleton, motor proteins and certain mechanosensitivies of actin-related proteins may implement comparison or contrast-based learning~\cite{banerjee2025learning, wang2025mechanosensitive}.

To use contrasts to autonomously adjust their tunable degrees of freedom, systems must implement ``local rules," \textit{i.e.}, they must change each tunable degree of freedom by comparing only partial information about the system in the two states~\cite{stern2023learning}. Local rules have been introduced for supervised learning~\cite{rumelhart1985learning,movellan1991contrastive,carreira2005contrastive, scellier2017equilibrium}, including in tunable non-biological physical systems that exploit physical processes to ``calculate" output responses to applied inputs.  The latter include flow networks  (\textit{in silico} \cite{stern2021supervised, stern2022physical,anisetti2023learning, anisetti2024frequency}) and electrical networks  (\textit{in silico}  \cite{kendall2020training};  laboratory~\cite{wycoff2022learning,dillavou2022demonstration, dillavou2024machine,stern2024training}) with tunable conductances, and mechanical networks with adjustable spring constants and/or equilibrium lengths (\textit{in silico} \cite{stern2020supervised}; laboratory \cite{arinze2023learning,altman2024experimental,li2024training}).
Contrastive learning has also been proposed for quantum systems~\cite{lopez2023self,wanjura2024quantum}, and implemented \textit{in silico} in Ising machines~\cite{laydevant2024training}. 
The laboratory systems in \cite{wycoff2022learning,dillavou2022demonstration, dillavou2024machine,stern2024training} do not require intervention by a processor or human; they adjust themselves autonomously.

In physical systems, contrastive learning has so far been restricted primarily to networks with tunable \textit{reciprocal} interactions in steady state or equilibrium. By ``reciprocal" interactions, we mean that the interaction between node $a$ and node $b$ is the same as the interaction between $b$ and $a$.  Conventional contrastive learning exploits the fact that a global scalar quantity (a Lyapunov function) such as energy, power, or action, is minimized in equilibrium or steady state. Recent treatments have extended contrastive learning to steady states in systems with tunable non-reciprocal interactions~\cite{scellier2018generalization} and to dynamical systems with tunable reciprocal Lagrangians~\cite{massar2025equilibrium,pourcel2025lagrangian,berneman2025equilibrium}.
However, living systems are typically not in steady state or equilibrium, and are also often active, driven by energy-injection at the microscopic scale~\cite{ramaswamy2010mechanics,needleman2017active,shaebani2020computational,shankar2022topological}, and have non-reciprocal interactions~\cite{fruchart2021non}. These properties can be helpful for learning -- recent studies of local learning in active matter~\cite{floyd2024learning,du2025metamaterials} and neuronal networks~\cite{fernandez2024ornstein,van2026dynamical, ellenberger2025backpropagation} show that some desired behaviors rely on features such as non-reciprocal information transfer between elements~\cite{scellier2018generalization, du2025metamaterials},  time-delayed information processing~\cite{falk2023learning,mandal2024learning}, and dissipation~\cite{stern2022physical,berneman2024designing,massar2025equilibrium}. 

Here, we extend the standard theory of contrastive learning to supervised learning in dynamical systems that are active, far from equilibrium or steady state, and/or that have non-reciprocal interactions. We develop local, physically realizable rules for training networks governed by coupled ordinary differential equations to follow desired dynamical trajectories. We first generalize a local equilibrium contrastive learning rule to reduce differences between  a ``free" trajectory where only input signals are applied, and a ``clamped" trajectory where a \emph{supervision protocol} nudges the output towards the correct outcome.  We then discuss the ``gradient supervisor," which guides the system to descend a cost function gradient via clamping. The clamped trajectory is determined from a local error signal, but, because we are working with dynamical systems, we must adjust the entire past behavior of the system to decrease instantaneous error in the present.  Thus, the gradient supervisor must clamp the trajectory of every node at every instant up to the present time to bring the system closer to the desired output. This procedure scales extensively with system size and is unworkable in practice. 

To address this challenge, we define Probably Approximately Right (PAR) supervision as a tractable alternative that is simple, local, and causal. These attributes come with the price of only approximating the cost gradient. Together, our local learning rule and supervision protocol directly generalize equilibrium/steady-state contrastive learning, and provide a recipe that applies to any dynamical system governed by coupled ordinary differential equations (ODEs). To demonstrate the broad applicability of our training process, we apply it successfully \textit{in silico} to tunable linear oscillator networks, Kuramoto  networks, leaky integrate-and-fire (LIF) neuronal networks, Michaelis-Menten chemical reaction networks, and generalized Lotka-Volterra models.

\section{Dynamical contrastive learning}
\label{sec:DynamicalContrastiveLearning}

Consider a \emph{tunable dynamical system}, a system of first-order coupled ordinary differential equations (ODEs),
\begin{equation}
\begin{aligned}
\dot{\vec{x}}^F = \vec{F}(\vec{x}^F(t;\vec{w}),\vec{w},t)
\end{aligned}
 \label{eq:ODE},
\end{equation}
where $\vec{x}(t;\vec{w})$ (components $x_a$) is
the trajectory over time $t$  of \emph{physical degrees of freedom}  (physical DOFs). We represent each system as a network with a different physical DOF at each node. The dynamics are controlled by $\vec{w}$,  \emph{tunable degrees of freedom} (tunable DOFs) defined on network edges indexed by $i$, and assumed fixed during a trajectory.  For supervised learning of input-output tasks, we take some nodes to be inputs where signals are applied, and some nodes to be outputs where responses are measured. The remaining nodes are referred to as ``hidden." The superscript $F$ denotes \emph{free trajectories} of the system in response to signals applied at input nodes, which we will compare later to \emph{clamped trajectories}, where supervisory signals are also applied to other nodes in the spirit of contrastive learning. The components of $\vec{F}$, $F_a$, are \emph{dynamical operators} determining total time derivatives $\dot{x}_a^F=\frac{d}{dt}x_a^F$.  Higher order ODEs (e.g., Newtonian dynamics) can be expanded to a larger system of first order ODEs, so Eq.~\eqref{eq:ODE} applies to any deterministic motion. The dynamical operators $F_a$ are local in space and time, so the physical DOF $x_a$ are only affected by instantaneous states of neighboring physical and tunable DOFs. Thus, the Jacobian matrices $\frac{\partial F_a}{\partial x_b}$ and $\frac{\partial F_a}{\partial w_i}$ will be sparse. The $F_a$, with additively included external inputs $I_a(t)$, act as ``forces" driving an overdamped system.

Given time-varying inputs $I_a(t)$ at input nodes, $\vec x_I$, and  initial conditions, $x_a^F(t=0) \equiv x_a^0$,  consider tasks where free trajectories minimize a trajectory-dependent cost 
\begin{equation}
C = \int_0^T dt \, c(\vec{x}^F(t;\vec{w})),
\label{eq:instantcost}
\end{equation}
where $c$ is an instantaneous cost function, and $T$ is the trajectory duration. Typically, the cost will be incurred only at output nodes $\vec x_O$ consisting of an $O(1)$ subset of the $N$ nodes. 

We first consider gradient descent on this cost function with respect to the tunable DOFs:
\begin{equation}
\Delta \vec{w}^G = -\alpha \vec{\nabla}_{w}C  \, ,  
\end{equation}
where $\alpha$  sets the learning rate.  
Formally, the components of the cost gradient are
\begin{equation}
\begin{aligned}
\frac{d C}{d w_i} &= \int_{0}^{T} dt \, \frac{d c(\vec{x}^F(t;\vec{w}))}{d w_i} =\int_{0}^{T} dt \, \frac{\partial c(\vec{x}^F)}{\partial x_a^F} \frac{d x^F_a(t;\vec{w})}{d w_i} \, .
\end{aligned}
 \label{eq:CG1}
\end{equation}
Here, and below, repeated indices are summed and we assume that $c$ depends on the tunable DOF $\vec{w}$ only through the physical DOF $\vec{x}^F$ (the extension to explicit  $\vec{w}$ dependence is in Appendix~\ref{sec:ExplicitDep}).

To compute the first term in Eq.~\eqref{eq:CG1}, $\frac{\partial c}{\partial x^F_a}$, which measures how much the instantaneous cost changes if the trajectory is deformed, we simply differentiate the cost function.  To compute the second factor in Eq.~\eqref{eq:CG1}, which measures how much the trajectory changes if we perturb the tunable DOFs, we employ a formal solution to Eq.~\eqref{eq:ODE}:
\begin{equation}
\begin{aligned}
x^F_a(t) &= x^0_a + \int_0^t dt' F_a(t')\\
\mathcal{J}_{ai}(t) &\equiv \frac{d x^F_a(t)}{d w_i}= \int_0^t dt' \,  \frac{d F_a (t^\prime)}{d w_i} \\
&= \int_0^t dt' \, \left[\frac{\partial F_a (t^\prime)}{\partial w_i} + \frac{\partial F_a (t^\prime)}{\partial x_b} \mathcal{J}_{bi}(t^\prime)\right]
\end{aligned}
 \label{eq:FreeTrajectory}
\end{equation}
where the last expression follows by differentiating the explicit and implicit dependencies on $\vec{w}$ in $F_a$.

The integral equation has a standard solution \cite{Schwartz2013QFT}:
\begin{equation}
\begin{aligned}
\mathcal{J}_{ai}(t) &= \int_0^t dt'  \, S_{ab}(t,t') \, \frac{\partial F_b(t')}{\partial w_i}  \\
S_{ab}(t,t')&\equiv\theta(t-t') \, \mathcal{T} \,\textrm{exp}\left(\int_{t'}^t dt'' \,
\frac{\partial F(t'')}{\partial x} \right)_{ab},
\end{aligned}
 \label{eq:Jacobian}
\end{equation}
where $\mathcal{T}$ is the time-ordering operator: $\mathcal{T}(A(t) B(t')) = \theta(t-t')A(t)B(t') + \theta(t'-t)B(t')A(t)$. The \emph{signal matrix} $S_{ab}(t,t')$ propagates effects of a change in the trajectory of $x_b$ at time $t'$ to a change in the trajectory of $x_a$ at  $t>t'$.  The step function $\theta(t-t')$ vanishes when $t<t'$, ensuring that  propagation is causal.  While $\frac{\partial F}{\partial x}$ is local, connecting nearby $x_{a,b}$, the matrix exponential spreads the effects throughout the network. Combined with Eq.~\eqref{eq:CG1}, the cost gradient is thus given
\begin{equation}
\begin{aligned}
{d C \over d w_i} &= \int_{0}^{T} dt \, \frac{\partial c(t)}{\partial x_a}\int_0^T dt'  \, S_{ab}(t,t')\, \frac{\partial F_b(t')}{\partial w_i} \, .
\end{aligned}
 \label{eq:CG2}
\end{equation}
Therefore, a rule that descends the cost gradient in Eq.~\eqref{eq:CG2} will be {\it nonlocal} in space and time, involving every node in the network. Hence, it cannot be realized through local physical interactions~\cite{stern2023learning}.

\subsection{A local rule}
\label{sec:LocalLearningRule}

{\it Local} contrastive learning rules have been used to train systems at equilibrium or at dynamical fixed points~\cite{movellan1991contrastive, scellier2017equilibrium, stern2021supervised,scellier2018generalization, scurria2026}.  In these approaches,  a {\it free state}, $\vec{x}^F$, determined by the response of the physical DOFs to inputs, is compared to a {\it clamped state}, $\vec{x}^C$, affected by a supervisor, to determine updates of 
tunable DOFs. The supervisor can either nudge the physical DOFs toward outputs that reduce cost (Equilibrium Propagation~\cite{scellier2017equilibrium}), or clamp them near the right place (Coupled Learning~\cite{stern2021supervised}).  Each tunable DOF locally compares free and clamped states.  In systems that minimize a scalar function, $E(\vec{x}^F)$, so that $\vec{F} = -\vec \nabla_{x} E$ in Eq.~\eqref{eq:ODE}, the local contrastive rule is 
\begin{equation}
  \Delta \vec{w} = -\alpha \vec \nabla_w [E(\vec{x}^C) - E(\vec{x}^F)] \,,
  \label{eq:localrule}
\end{equation} 
where $E$ is evaluated at its minimum for the free and clamped states. Contrastive learning, in the form of Equilibrium Propagation, follows the cost gradient in the limit of infinitesimal forcing by the supervisor~\cite{scellier2017equilibrium}. However,  we are interested here in systems that are not in equilibrium or at a steady state, and that may have non-reciprocal interactions, be non-conservative, or be subject to external dynamical driving. Such systems do not generally minimize any Lyapunov function $E$, so standard contrastive learning methods do not apply. 

Nevertheless, we (or a tunable DOF) can still locally compare free and clamped trajectories.  We model a \emph{clamped trajectory} as a slightly nudged free trajectory (Fig.~1),
\begin{equation}
\begin{aligned}
\vec{x}^C(t) \equiv \vec{x}^F(t) + \eta \, \vec{n}(t)
\end{aligned}
 \label{eq:ClampedTrajectory},
\end{equation}
where $\vec{n}(t)$ acts on some physical DOFs. We take  $\eta\ll 1$ so that the nudge acts linearly around the free trajectory, and start by considering conservative systems so that $\vec{F} = - \vec \nabla_x E$ in (\ref{eq:ODE}).  
 
To devise a local rule we expand the partial derivative of the clamped energy with respect to the tunable DOFs: 
\begin{equation}
\begin{aligned}
\vec \nabla_w{E(\vec{x}^C)}  &= \vec \nabla_w (E(\vec{x}^F+(\vec{x}^C-\vec{x}^F))\\
&\approx \vec \nabla_w E(\vec{x}^F) + (\vec{x}^C-\vec{x}^F) \cdot \vec \nabla_w \vec \nabla_x E(\vec{x}^F) \\
&= \vec \nabla_w E(\vec{x}^F) - (\vec{x}^C-\vec{x}^F) \cdot \vec \nabla_w \vec{F}(\vec x^F)
\end{aligned}
 \label{eq:GC}
\end{equation}
where we  assumed $\vec{x}^C-\vec{x}^F\sim O(\eta)$. Using this expression in the contrastive rule, Eq.~\eqref{eq:localrule}, gives~\cite{stern2024physical}:
\begin{equation}
\begin{aligned}
\Delta w_i  &= -\frac{\alpha}{\eta}\frac{\partial}{\partial {w_i}} [E(\vec{x}^C) - E(\vec{x}^F)]\\
&=\frac{\alpha}{\eta}(\vec{x}^C-\vec{x}^F) \cdot \frac{\partial \vec{F}}{\partial w_i} .
\end{aligned}
 \label{eq:LR1}
\end{equation}

\begin{figure}
\includegraphics[width=1.0\linewidth]{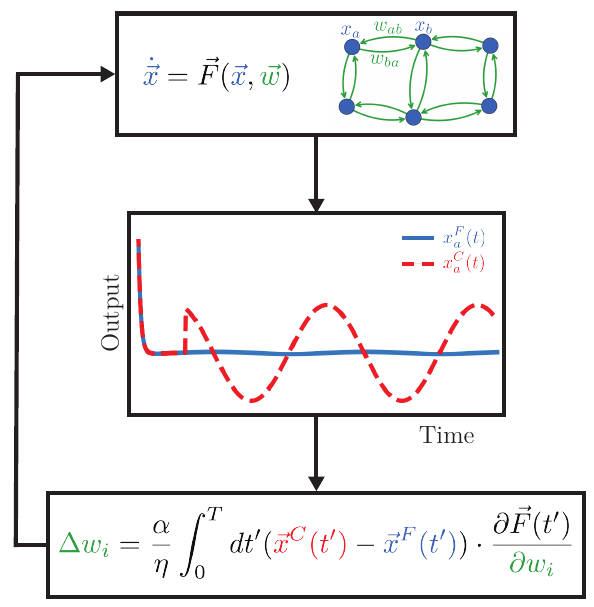}
\caption
{Dynamical contrastive learning. We consider dynamical systems in which physical variables $x_a(t)$, indexed by $a$, are coupled through learning degrees of freedom, $w_{ab}$.  The effect of $x_a(t)$ on $x_b(t)$, set by $w_{ab}$, may be different from the effect of $x_b(t)$ on $x_a(t)$, set by $w_{ba}$. The system learns by adapting $\vec{w}$ based on comparison of a free trajectory $\vec x^F(t)$ and a clamped trajectory $\vec x^C(t)$ nudged by an external supervisor (Eq.~\eqref{eq:LR3}). During learning, the free trajectory iteratively approaches the clamped trajectory, reducing a cost function.
}
\label{fig:figSketch}
\end{figure}

This is a network-local contrastive rule that applies to conservative systems in equilibrium or at steady state~\cite{stern2024physical}.  Scellier et. al.~\cite{scellier2018generalization} used a similar expression to define a local rule in non-conservative systems settling at fixed points. To generalize to dynamical trajectories, we  consider free and clamped trajectories $\vec{x}^F(t)$ and $\vec{x}^C(t)$ that depend on the force $\vec{F}(t)$ which can be non-conservative. Suppose we apply the rule Eq.~\eqref{eq:LR1} instantaneously at each moment in time: over a duration $T$, the integrated change to the tunable DOF (Fig.~\ref{fig:figSketch}) is 
\begin{equation}
\begin{aligned}
\Delta w_i^{\text{tot}} = \frac{\alpha}{\eta} \int_0^T  dt'  \, (\vec{x}^C(t') - \vec{x}^F(t'))
\cdot \frac{\partial \vec{F}(t')}{\partial w_i}
\end{aligned}
 \label{eq:LR3}.
\end{equation}
This is the contrastive local rule that we will use.
Here $\Delta W = -(\vec{x}^C-\vec{x}^F) \cdot \vec{F}$ is the infinitesimal work performed by nudging the free trajectory to the clamped one.  Thus,  Eq.~\eqref{eq:LR3}  minimizes the \emph{integrated work needed to bring the free trajectory to the clamped one}, a well-defined notion for non-conservative as well as conservative systems.  

\subsection{The gradient supervisor}
The rule in Eq.~\eqref{eq:LR3} is  local in space and can be determined from local measurements in time, given a clamped trajectory $\vec{x}^C$. But how should the clamped trajectory, $\vec{x}^C(t)$, be chosen?   Ideally, the local rule in Eq.~\eqref{eq:LR3} should descend the gradient of the cost, Eq.~\eqref{eq:CG2}, so that 
\begin{equation}
\begin{aligned}
\Delta w_i & =-\alpha {d C \over d w_i} = - \alpha \int_0^T dt' \, \vec{r}(t') \cdot {\partial \vec{F}(t') \over \partial w_i} \\
r_b(t') &= \int_0^T dt \, {\partial c(\vec{x}^F(t)) \over \partial x_a} \,S_{ab}(t,t') \, .
\label{eq:GradientDescentRule}
\end{aligned}
\end{equation}
where we exchanged the order of integrals in Eq.~\eqref{eq:CG2} and evaluated the inner integral to write a trajectory $\vec{r}(t')$. Now, if the supervisor chooses the clamped trajectory as
\begin{equation}
\begin{aligned}
\vec{x}^C(t')= \vec{x}^F(t')- \eta \, \vec{r}(t'), \, 
\end{aligned}
 \label{eq:ClampedTrajectory2} 
\end{equation}
then the local rule in Eq.~\eqref{eq:LR3} performs gradient descent, Eq.~\eqref{eq:GradientDescentRule}. We will refer to Eq.~\eqref{eq:ClampedTrajectory2} as the ``gradient supervisor" because it implements exact global gradient descent in the limit of infinitesimal nudges.  

The easy part of implementing the gradient supervisor is to observe the free trajectory and compute the cost derivative  $\frac{\partial c}{\partial x_a}$ prior to clamping at the subsequent iteration. The challenging part is the nonlocal computation of the signal matrix $S_{ab}(t', t)$ and the integral $\int dt \, {\partial c(t)\over \partial x_a}S_{ab}(t,t')$ to determine the nudged trajectory $\vec{r}(t')$. This procedure is necessary because, given enough time, the error signal propagates across the entire system. Thus, in order to descend the cost gradient the supervisor must nudge \emph{every} node at each step, not just the output nodes.  Moreover, the supervisor must calculate the nudge at each node, a system-wide calculation.  As a result, the gradient supervisor requires computation that scales with system size, rendering it impossible to implement in sufficiently large systems, even via offline computation. In other words, the learning rule is local, but the gradient supervisor is not.

The contribution of the integrand defining $\vec{r}(t')$ in Eq.~\eqref{eq:GradientDescentRule}, which determines the clamped trajectory via Eq.~\eqref{eq:ClampedTrajectory2}, is the dynamical solution of a system initialized at time $t$ with $x_a^C(t)=x^F_a(t)-\eta \frac{\partial c(t)}{\partial x_a}$ and propagated \emph{backwards} to  $t'$ under the influence of an appropriately time-reversed dynamical operator $\bar{F}_a$. This means that a supervisor who clamps only output nodes at times $t$ could yield the exact cost gradient if the system could propagate the error signal backward in time, accumulating the propagated error~\cite{pourcel2025lagrangian,pourcel2025learning}. 

If the system has reciprocal interactions and time-reversal invariance, forward and backward propagation in time are identical, in which case the gradient supervisor can be implemented through causal propagation. But general systems with non-reciprocal interactions break time-reversal symmetry. As a result, exact gradient computation with a local rule requires either a supervisor who computes the non-local signal propagator $S_{ab}(t,t')$ and clamps dynamical trajectories of every single node accordingly, or a supervisor who clamps \emph{only} output nodes, but propagates the error signal backwards in time to every node. 

\subsection{The forward supervisor}

We seek a supervisor who measures local error only at the output nodes $x_O$, clamps only output nodes, and relies on physics to propagate the supervision through the rest of the network. Since physics is causal, information about an error at time $t'$ can only propagate to later times, $t>t'$. Clearly, this supervisor cannot reproduce exact gradient descent.

Suppose our supervisor nudges the clamped trajectory by infinitesimally pushing output nodes, $\vec{x}^C(t')=\vec{x}^F(t')+\eta \vec{n}(t')$, at a time $t'$ before which the free and clamped trajectories coincide. Setting aside what the nudge should be, we can ask how the clamped trajectory of the  system evolves in response. First, we take the derivative of the free trajectory in Eq.~\eqref{eq:FreeTrajectory} with respect to the physical degrees of freedom and obtain for $t>t^\prime$
\begin{equation}
\begin{aligned}
\mathcal{K}_{ab}(t,t') &\equiv \frac{d x_a^F(t)}{d x_b(t')} \\
&=
\delta_{ab} + \theta(t-t')\int_{t'}^t dt'' \, \frac{\partial F_a}{\partial x_c} (t'')\mathcal{K}_{cb}(t'',t').
\end{aligned}
 \label{eq:FreeTrajectory2}
\end{equation}
To arrive at this self-consistency equation, we write $x_a^F(t)$ as the time derivative of the formal solution in Eq.~\eqref{eq:FreeTrajectory}, and manipulate the resulting expression. The first term accounts for a nudge in the initial condition at node $b$. The second term implies that changes in any $x_c$ at any time modify the dynamical operator $\partial F_a / \partial x_c$, and this change propagates self-consistently forward in time.

We can again formally solve this Volterra integral equation as a power series, yielding,
\begin{equation}
\begin{aligned}
\mathcal{K}_{ab}(t,t') =\theta(t-t') \, \mathcal{T} \,\textrm{exp}\left(\int_{t'}^t dt''  \,\frac{\partial F}{\partial x} \right)_{ab} = S_{ab}(t,t')
\end{aligned}.
 \label{eq:NudgePropagator}
\end{equation}
In other words, the nudges of the clamped trajectory are propagated by the signal matrix. For an infinitesimal nudge $\eta \vec{n}(t')$ at time $t'$, the trajectory at later times changes by $\vec{x}^C(t)-\vec{x}^F(t)\approx \eta S(t,t') \vec{n}(t') $. If a supervisor continually applies infinitesimal nudges to the clamped trajectory during  dynamics, there is a cumulative effect:
\begin{equation}
\begin{aligned}
\vec{x}^C(t) \approx \vec{x}^F(t) + \eta \int_0^T dt' \, S_{ab}(t,t') n_b(t')
\end{aligned}.
 \label{eq:NudgedDynamics}
\end{equation}
where the repeated index is summed. 

Note that we do not account for the effects of accumulating intermediate changes in the trajectory due to previous nudging, as this effect scales with $\eta^2$. Using these dynamics in the local learning rule Eq.~\eqref{eq:LR3}, we have
\begin{equation}
\begin{aligned}
\Delta w_i = \alpha \int_0^T dt \, \bigg[\int_0^T dt' \, S_{ab}(t,t') n_b(t') \,  \bigg]
\frac{\partial F_a (t)}{\partial w_i}
\, .
\end{aligned}
 \label{eq:NewSup}
\end{equation}

Finally, we can consider a  supervisor who only nudges output nodes when an error is measured, 
\begin{equation}
\begin{aligned}
n_O(t')=- \frac{\partial c(t')}{\partial x_O} .
\end{aligned}
 \label{eq:NewSupX}
\end{equation}
Combined with the local learning rule, the change in the tunable DOFs is
\begin{equation}
\begin{aligned}
\Delta w_i = -\alpha \int_0^T dt \, \bigg[\int_0^T dt' \, S_{aO}(t',t) \frac{\partial c(t')}{\partial x_O} \,  \bigg]
\frac{\partial F_a (t)}{\partial w_{i}}
\, .
\end{aligned}
 \label{eq:NewSup2}
\end{equation}
This is what we call the ``forward supervisor." There are similarities between Eq.~\eqref{eq:NewSup2} and the gradient supervisor computed in Eq.~\eqref{eq:GradientDescentRule}, but note that the node indices and the ordering of $t$ and $t^\prime$ are reversed. The time reversal occurs because the error signal is propagated forward in time rather than backward. The forward supervisor relies on causal propagation of the error signal, unlike the gradient supervisor, which propagates the signal backwards in time. The reversal of node indices transposes the signal matrix $S$ relative to the one in Eq.~\eqref{eq:GradientDescentRule}. As a result, our tractable supervisor is different from the gradient supervisor, and implementation of the local rule with this supervisor does not lead to gradient descent on the cost in any limit. However, we will see that successful learning still follows if Eq.~\eqref{eq:GradientDescentRule} and Eq.~\eqref{eq:NewSup2} are positively correlated.

\subsection{Probably Approximately Right (PAR) supervision}

We have demonstrated that local rules with tractable supervisors cannot recover gradient descent on the cost when the dynamics violate time-reversal symmetry. The gradient supervisor (Eq.~\eqref{eq:GradientDescentRule}) explicitly contains global information about what happened in the system before the error occurs, while our proposed supervisor can only affect the system \emph{after} the error has occurred because a causal supervisor cannot backpropagate error in time. A gradient supervisor is, in principle, tractable for systems with reciprocal edge interactions in time reversal symmetric~\cite{lopez2023self,massar2025equilibrium,pourcel2025learning} or time-periodic systems~\cite{berneman2025equilibrium}. 
However, supervisors implemented in real, laboratory systems can never exactly recover gradient descent even if reciprocal, because they cannot apply infinitesimal nudges.  Additionally, any real physical system will contain imperfections, such as noise sources and biases in implementing local rules. Such imperfections can lead to dynamical behavior that breaks time reversal symmetry even in nominally reciprocal systems trained for equilibrium tasks~\cite{dillavou2025understanding}. 

Despite this challenge, we expect that tunable dynamical systems will learn successfully if instantaneous adjustments to each tunable DOF during training are at least approximately aligned with the cost gradient.  In that case, each training step will reduce the cost, albeit more slowly than in exact gradient descent.  In fact, the training steps do not even have to be correlated with the gradient at all times -- they can sometimes have negative projection onto the gradient, provided they move the system in the correct direction most of the time, or at least on average.  In other words we expect only to need Probably Approximately Right (PAR) supervision. (We adopt this terminology in homage to Probably Approximately Correct (PAC) learning in the machine learning literature~\cite{valiant1984theory}, but avoid the latter term because of its many existing associations, which do not apply here.) For example, the correlation between the cost gradient and its physical local approximation could fluctuate between being positive and negative, provided it is positive often enough, especially as the system approaches a solution to the desired task.  Many forms of PAR supervision can be envisioned, including strong ones in which the correlation is required to be positive with high probability, but here we consider a weak version in which we simply require a positive correlation with the gradient on average:
\begin{equation}
\begin{aligned}
\langle\Delta w^{\text{Gradient}}_i \Delta w^{\text{Local}}_i \rangle > 0 .
\end{aligned}
 \label{eq:WeakCorr}
\end{equation}
In this view, the exact gradient is not generally implementable, but is simply a useful benchmark for local rule approximations to the gradient.  For different systems and tasks, different combinations of local learning rules and PAR supervision protocols may be needed to maximize average correlation in Eq.~\eqref{eq:WeakCorr}. In other words, we propose that focus in the field should shift from trying to recover the exact gradient of the cost to devising better tractable learning rules and PAR supervisors, consistent with the physical system and task of interest~\cite{dillavou2025understanding, ezraty2025harnessing}.

Average positive alignment between our local rule and forward supervisor and the  gradient is formally guaranteed in certain situations. For a simple short-time task involving one input and one output node, we show in Appendix B that 
\begin{equation}
\begin{aligned}
\frac{\Delta w^{\text{Gradient}}_i \Delta w^{\text{Local}}_i}{\vert \Delta w^{\text{Gradient}}_i \vert \vert \Delta w^{\text{Local}}_i\vert} =  1-\mathcal{O}\left(T t^*\right) , 
\end{aligned}
 \label{eq:AlignmentFirstOrder}
\end{equation}
where $t^*$ represents the duration from the initial condition until the time the error occurs, and $T$ denotes the characteristic memory or signal decay time of the system.
In the short-time limit, where the supervisor acts soon after the initial condition, and the local rule only accounts for short times after the supervisor's intervention, this framework effectively recovers the gradient to first order. However, while this proof identifies a mathematically tractable case where our learning framework is guaranteed to succeed, this scenario is quite restrictive and does not correspond to typical cases of interest in which systems are trained for temporal tasks that extend far beyond the initial condition. 

While we cannot \emph{prove} that our proposed forward supervisor satisfies the weak PAR criterion of Eq.~\eqref{eq:WeakCorr}, we will demonstrate numerically in the remainder of this paper that the condition Eq.~\eqref{eq:WeakCorr} is nevertheless met by our combination of local contrastive rule and forward supervisor for various linear and non-linear dynamical systems tuned to perform tasks far from the regime of the proof in Appendix B.  It would be useful to develop a more general theory of when PAR supervision, defined either by the average condition Eq.~\eqref{eq:WeakCorr} or in some other way, applies and leads to successful learning.

\section{Demonstrations}
\label{sec:demonstrations}

Below we apply the local learning rule of Eq.~\eqref{eq:LR3} with the forward supervisor of  Eq.~\eqref{eq:NewSup2} to five tunable systems: (A) Networks of coupled linear oscillators, (B) Kuramoto oscillator networks, (C) Leaky Integrate-and-Fire (LIF) neuron networks, (D) Michaelis-Menten chemical reaction networks, and (E) generalized Lotka-Volterra models.

\subsection{Coupled linear oscillators}
\label{sec:CoupledLinear}
Consider a tunable network of $N$ coupled linear oscillators:
\begin{equation}
\dot{x}_a = \sum_{a\neq b} ^N A_{ab} x_b,
\label{eq:linear_dynamics}
\end{equation}
where $x_a$ is the position of the $a$th oscillator, $A_{ab}$ and $A_{ba}$ are the tunable DOFs coupling oscillators $a$ and $b$.

For our first task,  consider a network of $N=10$ fully-connected, {\it reciprocally}-coupled oscillators with $A_{ab}=A_{ba}$ in Eq.~\eqref{eq:linear_dynamics}. One oscillator is the input, one is the output, and the remainder are ``hidden'' nodes.  We train for the {\it amplitude task} where the time-dependent output displacement mirrors the input displacement but amplified by a factor $p$.

\begin{figure}
\includegraphics[width=1.0\linewidth]{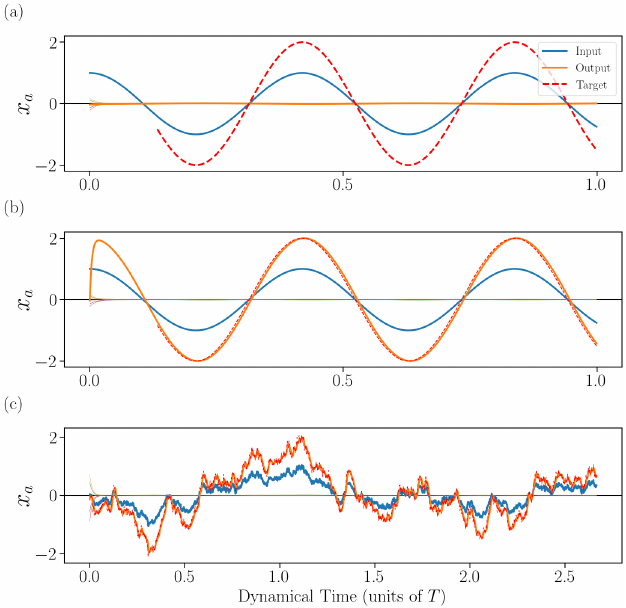}
\caption{Training reciprocal linear dynamical systems for an output response proportional by a factor $p$ to the input signal. ({\bf a}) The desired output signal (dashed red line) has double the amplitude ($p=2$) of the sinusoidal input signal at the input node, $x_1$ (solid blue line). Before training, the output response $x_{10}$ (orange) in the initialized network is weak. During training, the output node is clamped with a nudge from the free output signal towards the desired signal. After one training period of the input signal, the weights are adjusted according to the local rule of Eq.~\eqref{eq:LR3} and the clamping on the output node is adjusted according to the forward supervisor of Eq.~\eqref{eq:NewSup2}.  This process is repeated until the desired output signal is achieved within acceptable error. This training paradigm is applied to each of the examples in the following figures unless otherwise noted in the main text. ({\bf b}) After training, the output oscillator (orange) matches the desired trajectory and the hidden oscillators have weak responses (thin lines near 0).
({\bf c}) After training, the network generalizes the amplitude-doubling  to random input data (blue): the output (orange) matches the desired output (red dashed) well.  
}
\label{fig:figLinearAmplitude}
\end{figure}

\begin{figure*}
\includegraphics[width=1.0\linewidth]{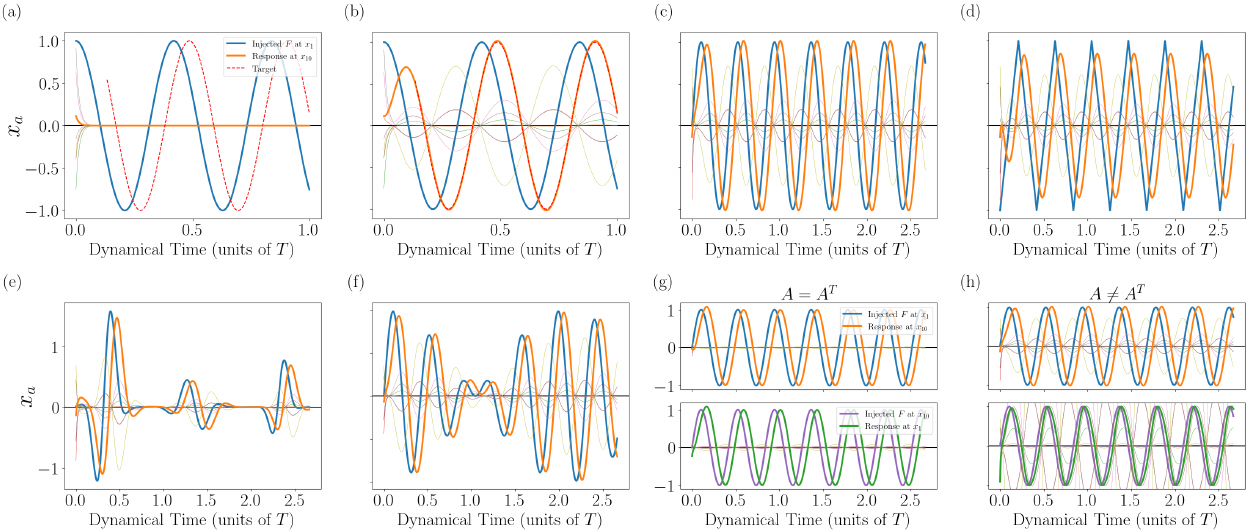}
\caption{Training a linear dynamical system to promote a desired temporal lag in the output. ({\bf a}) The desired output signal (dashed red line) lags the input signal, $F$ (blue) by a time interval. Before training, the response of the output node, $x_{10}$ (orange), and hidden nodes (thin colored lines) in the initialized network are weak ({\bf b}) After training, the output node reproduces the desired trajectory. ({\bf c}) After training is complete, the network generalizes to a longer time window for an input sinusoid at a different initial phase and for random initial node positions. The network also generalizes correctly for input ({\bf d}) triangle waves, ({\bf e}) impulse signals, and ({\bf f}) composite waves. ({\bf g}) (Top) Response of ``output node," $x_{10}$ (orange) when a sinusoid is injected at ``input node" $x_1$, (blue) vs. (Bottom) response of previous ``input node" $x_1$ (green) when a sinusoid is instead injected at the previous ``output node" $x_{10}$ (purple), showing reciprocal time lags in a trained reciprocal system. ({\bf h}) Same as (g) but for a non-reciprocal network trained for two different time lags, depending on whether a signal is injected at one node or the other.}
\label{fig:figLinearLag}
\end{figure*}

Although we train this system using a dynamical local rule, Eq.~\eqref{eq:linear_update_symmetric} and the forward supervisor in Eq.~\eqref{eq:NewSup2}, the task is in fact static: the input-output relation simply requires that $x_\text{out}(t) = p x_\text{in}(t)$. Further, since the $A_{ab}$ are reciprocal, this network minimizes a global Lyapunov function and could be trained by static local rules and supervisors, as in  Equilibrium Propagation~\cite{scellier2017equilibrium} or Coupled Learning~\cite{stern2021supervised}. This first example is therefore designed to demonstrate that our local rule and supervisor can be applied to dynamical trajectories to learn static tasks. 

In the free state, we only impose the input trajectory, while in the clamped state, we impose the input trajectory while clamping the output oscillator towards the desired trajectory: $x_{\text{output}}^C(t) =x_{\text{output}}^F(t) (1-\eta)+\eta x_{\text{output}}^\text{target}(t)$ with a small nudging parameter $\eta\ll 1$. The hidden nodes respond to these input-output signals.  The update rule from Eq.~\eqref{eq:LR3} is 
\begin{multline}
    \dot{A}_{ab} = \frac{1}{2}  \frac{\alpha}{\eta}  \int_0^T dt  [ (x_a^C(t)-x_a^F(t)) x_b^F(t) \\
    + (x_a^C(t)-x_a^F(t)) x_a^F(t)]
    \label{eq:linear_update_symmetric}
\end{multline}
where $\alpha$ is a learning rate and $T$ is the duration of the desired trajectory. As noted earlier, the learning rule here, Eq.~\eqref{eq:linear_update_symmetric}, is symmetrized compared to Eq.~\eqref{eq:LR3} so that $A_{ab}$ are guaranteed to remain reciprocal.

With this rule, we train the system iteratively. We expose the system to a small number of periods of a sinusoidal wave at the input oscillator, $x_1$, and then nudge the output oscillator, $x_{10}$ towards an in-phase sinusoid with a different amplitude (Fig.~\ref{fig:figLinearAmplitude}a). After training, the output trajectory mirrors the target (Fig.~\ref{fig:figLinearAmplitude}b). To test generalization, we input a random signal over a longer time period (Fig.~\ref{fig:figLinearAmplitude}c), which the output signal should double (red dashed), as is indeed observed (orange).

Next, consider a {\it lag task} with $N=10$ fully-connected oscillators, now allowing non-reciprocity $A_{ab} \ne A_{ba}$. For this network, we inject a signal, $CF_{\text{input}}(t)$ for a constant $C=5$, locally at an input oscillator and measure the response at a different output oscillator. We then train the system, requiring the output displacement, $x_{\text{output}}(t)$, to have the same form as $F_\text{input}(t)$, but with a specified time lag $\Delta t$. Unlike the previous task, this one is intrinsically dynamical, requiring memory of the past state of the network to be encoded in the learned dynamics. Following Eq.~\eqref{eq:LR3}, the update rule is 
\begin{equation}
    \dot{A}_{ab} = \frac{\alpha}{\eta} \int_0^\tau dt [ x_a^C(t)-x_a^F(t) ]x_b^F(t). 
    \label{eq:linear_update_nonsymmetric}
\end{equation}
We again train the network by repeatedly presenting a small number of periods of an oscillatory wave at frequency $\omega_\text{train}$, and iteratively nudging the clamped state toward the desired temporal lag. Fig.~\ref{fig:figLinearLag}a shows the network before and after training. Although trained on a sinusoid applied to the input node, the network generalizes the temporal response to diverse dynamical input signals applied to that node, successfully producing the same time lag for each signal (Figs.~\ref{fig:figLinearLag}b-f).  

We used a non-reciprocal network above, yet a reciprocal network can learn the task too.  However, for a linear system with reciprocal interactions, certain restrictions do apply~\cite{Willems1972Reciprocal}. Denote node $A$ as the input and node $B$ as the output. Suppose node $B$ produces a displacement that mirrors the input at node $A$ with a time lag $\Delta t$. If an input is then applied to node $B$ in such a trained reciprocal network, the displacement signal at node $A$ must reproduce the input signal with an \emph{equivalent} time lag $\Delta t$ (Fig.~\ref{fig:figLinearLag}g). A non-reciprocal network, on the other hand, can be trained to have time lag $\Delta t$ in the displacement at node $B$ when the input is applied to node $A$, and an arbitrarily different time lag $\Delta t'$ in the displacement at $A$ when the input is applied to node $B$. Our learning protocol successfully trains the network to perform this more complex task as well (Fig.~\ref{fig:figLinearLag}h),  demonstrating a case that requires learning with dynamical, non-reciprocal elements in linear systems.

\begin{figure}
\includegraphics[width=1.0\linewidth]{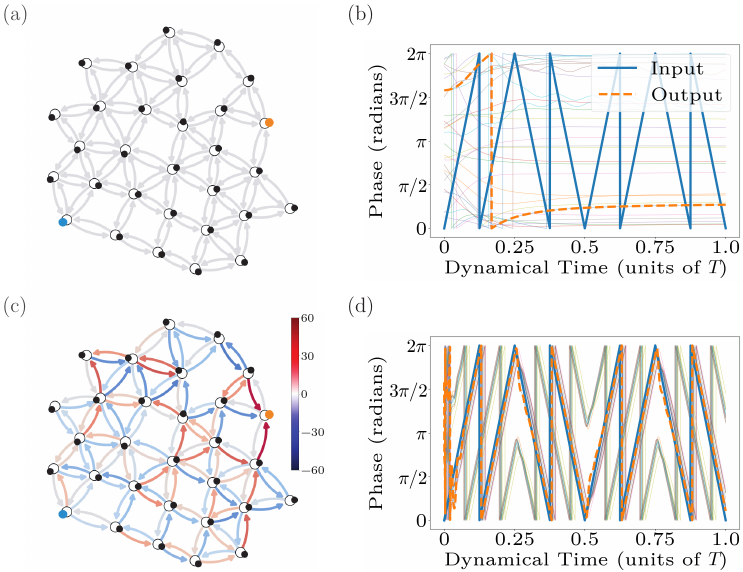}
\caption{Training a network of Kuramoto oscillators so that the output oscillator mirrors the dynamics applied at the input oscillator. ({\bf a}) Before training, the network has weak couplings between nearest neighbors. We want to transmit a signal from an input oscillator (blue) to on output oscillator (orange) on the opposite side of the network. ({\bf b}) Before training, a time-varying signal at the input oscillator (blue), produces a weak response at the target (dashed orange). The responses of nearly all oscillators in the network are also weak (thin lines). ({\bf c}) After training, some network couplings strengthen or weaken significantly, indicated by edge colors. ({\bf d}) The trained network achieves a dynamical response at the output (dashed orange) that is identical to the input signal (blue solid), as desired, by synchronizing nodes throughout the network.
}
\label{fig:figKuramoto_allo}
\end{figure}

\subsection{Kuramoto network}
\label{sec:Kuramoto}

\begin{figure*}[t]
\includegraphics[width=0.85\linewidth]{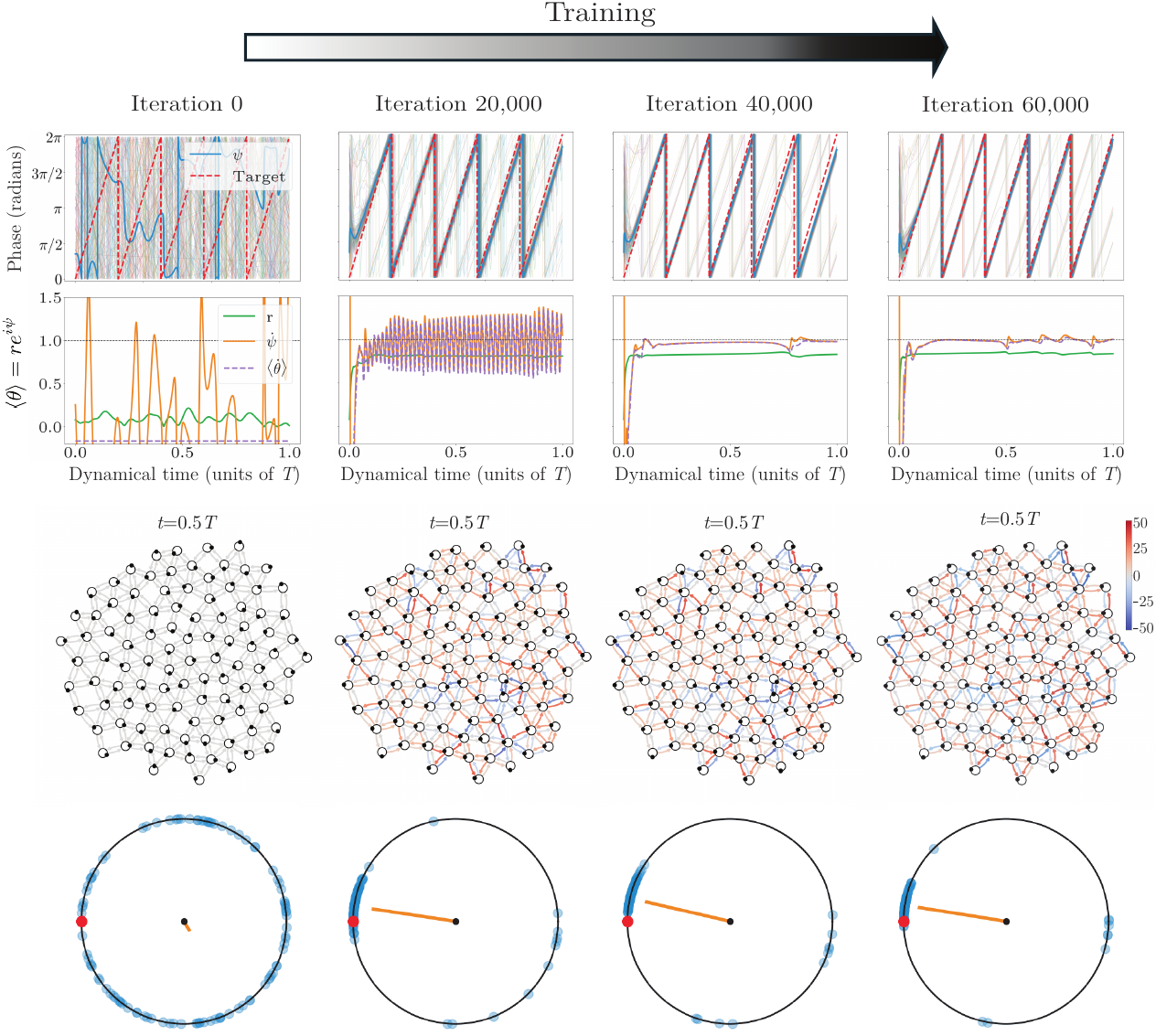}
\caption{Training snapshots for a network of $N=100$ Kuramoto oscillators spatially organized in two dimensions and coupled locally with an average of $Z=4.62$ neighbors.  The oscillators have intrinsic frequencies $\omega_a$ drawn from a normal distribution with mean zero and unit variance and are initially distributed uniformly in phase with vanishing couplings. The desired behavior is for the oscillators to synchronize to a common global frequency, here $\Omega = \dot{\psi} = 1$ so the desired period of oscillation is $2\pi$; the desired phase for each oscillator as a function of time is shown for five periods (the duration of the trajectory, $T$) by the red dashed line in each plot in the top row. Early in training (left column) no synchronization occurs because the oscillators are decoupled and none of them has the desired phase behavior. As training progresses (second, third and fourth columns), the couplings adjust so that nearly all of the oscillators acquire the desired behavior. 
({\bf Top row}) Desired phase (red dashed), actual phase of the Kuramoto order parameter $\psi(t)$ (Eq.~\eqref{eq:Kuramotoorderparameter}) (blue), and phases of individual oscillators (faded colors), over five periods of the desired oscillation, after the number of training periods, each of duration $T$, indicated by the top labels. At the end of training the oscillators have synchronized to the desired frequency. 
({\bf Second row})  Magnitude ($r$) and frequency ($\dot{\psi}$) of the Kuramoto order parameter over time during the same time intervals depicted in the top row.  The magnitude, $r$, grows with training and the frequency approaches the desired value $\Omega \equiv \dot{\psi} \rightarrow 1 = \omega_\text{sync}$, indicating synchrony at the desired frequency. 
({\bf Third row}) Local coupling strengths $K_{ab}$ (blue = negative, red = positive) between oscillators (circles), indicated at a time $t=T/2$ within each training iteration.  The couplings are non-reciprocal with some growing over time as the oscillators synchronize (phases at $t=T/2$ indicated as a black dot in the circle representing each oscillator). 
({\bf Fourth row}) Phases of all oscillators (blue dot), target phase (red dot),
and Kuramoto order parameter $re^{i\psi}$ (orange line, length = r, angle = $\psi$) at a time $t=T/2$ within each training iteration, showing that nearly all of them are synchronized by the end of training.
}
\label{fig:figKuramoto}
\end{figure*}

Next we consider the Kuramoto model describing the nonlinear dynamics of $N$ coupled phase oscillators
\cite{acebron2005kuramoto}:
\begin{equation}
    \dot{\theta}_a = \omega_a + \sum_{b=1}^{N} K_{ab}\sin(\theta_b-\theta_a)
    \label{eq:Kuramoto_dyn}
\end{equation}
where $\theta_a$ is the phase of the $a$th oscillator, $\omega_a$ is its intrinsic frequency, and $K_{ab}$ is the coupling between oscillator $a$ and $b$, which can be different from $K_{ba}$, the coupling between $b$ and $a$. Note that oscillators in an all-to-all Kuramoto network with uniform (fixed) coupling  $K_{ab} = K \sim \mathcal{O}(1/N)$ and a sufficiently narrow distribution of intrinsic frequencies will synchronize to a single global frequency \cite{Kuramoto1979Kuramoto} near the mean of the individual frequencies, a result that extends to other settings \cite{Kuramoto1984Chemical,acebron2005kuramoto}.  

Following Eq.~\eqref{eq:LR3}, the update rule for $K_{ab}$ is 
\begin{multline}
\dot{K}_{ab} = \frac{\alpha}{\eta} \int_0^T dt [(\theta_a^C - \theta_a^F) \sin(\theta_b^F - \theta_a^F)]\\
 \approx \frac{\alpha}{\eta}  \int_0^T dt [\sin(\theta_a^C - \theta_a^F) \sin(\theta_b^F - \theta_a^F)],
 \label{eq:lrkuramoto}
\end{multline}
where the second equality is at leading order in $\eta\ll 1$.

First, we consider learning an  {\it allostery task}~\cite{rocks2017designing}, in which the phases of a single input and output node should be matched. We train on a network of $N=30$ phase oscillators each with $\omega_a = 0$ and couplings initialized to zero. The oscillators are coupled sparsely to nearest neighbors with unidirectional connections and a neighbor distribution taken from a jammed packing of disks, with $\langle Z \rangle=4.1$ coupled neighbors on average per oscillator. We select one oscillator as an input and another distant one in the network as the output (Fig.~\ref{fig:figKuramoto_allo}a). We train the output oscillator to follow the distant input oscillator by clamping the output trajectory to mirror the input. During training, we repeatedly present the input-output pair with the training trajectory of a triangle wave at the source, corresponding to an alternating clockwise and counterclockwise rotation of the Kuramoto oscillator. This alternation is repeated $10,000$ times (epochs). Before training, the response at the output is weak (Fig.~\ref{fig:figKuramoto_allo}b). However, over the course of training, the bonds linking the input oscillator and output oscillator are strengthened (Fig.~\ref{fig:figKuramoto_allo}c) and the output oscillator exactly follows the input trajectory (Fig.~\ref{fig:figKuramoto_allo}d).  These results mirror those for the coupled linear system, though here we train in a non-reciprocal, non-linear system.

Next,  consider $N=100$ Kuramoto oscillators with nonzero intrinsic frequencies sampled independently from $\omega_{a}\sim\mathcal{N}(0,1)$, where $\mathcal{N}(\mu,\sigma^2)$ is a mean $\mu$ normal distribution with standard deviation $\sigma$. The coupling $K_{ab}$ are initialized to zero. The oscillators are again coupled sparsely to nearest neighbors, on average $\langle Z \rangle=4.62$  neighbors per oscillator. We train the network to synchronize globally at a frequency $\omega_{\text{sync}} = 1$ by updating the coupling matrix $K_{ab}$, while the intrinsic frequencies are fixed. 
During training of this {\it synchronization task}, in the clamped state, we nudge \textit{all} oscillators towards the desired trajectory $\theta_i^{\text{target}} = \cos(\omega_\text{sync} t)$. To evaluate the effect of training, we define the order parameter \cite{acebron2005kuramoto}:
\begin{equation}
    re^{i\psi} \equiv \langle e^{i\theta} \rangle 
    \label{eq:Kuramotoorderparameter}
\end{equation}
where $\langle \cdot \rangle $ indicates an ensemble average over all oscillators, and $r \in [0,1]$ indicates the degree of synchrony, with a fully phase-coherent state yielding $r\rightarrow 1$ and a completely phase-incoherent state yielding $r\rightarrow 0$. The phase of the synchronized system is $\psi$.

We train over 60,000 iterations of duration $T = 5 \times 2\pi/\omega_\text{sync}$. In each iteration, we clamp according to the forward supervisor, Eq.~\eqref{eq:NewSup2} over the time $T$. We compare to the free system over the same duration, updating the tunable degrees of freedom according to Eq.~\eqref{eq:LR3}, which for this system is Eq.~\eqref{eq:lrkuramoto}.  At the beginning of each iteration, we initialize the oscillators at set  positions, $\theta_a(0)$, selected from a uniform distribution $\theta_a(0) \sim U(0,2\pi)$. Over training, the Kuramoto order parameter magnitude $r$ grows and  $\dot{\psi} \rightarrow 1 = \omega_\text{sync}$, indicating synchronization at the desired frequency (Fig.~\ref{fig:figKuramoto}, second row). The magnitude of the nonsymmetric couplings grows as well, as nearly all the oscillators converge toward the same phase (Fig.~\ref{fig:figKuramoto}, third and fourth rows).

In order to achieve the desired global response in this task, the system \textit{requires} non-reciprocal connections --
reciprocal adaptive networks fail to achieve the desired synchronized frequency (see Appendix~\ref{sec:recipKuramoto}). This can be understood as follows: Kuramoto oscillator networks tend to synchronize robustly~\cite{acebron2005kuramoto}, but only near the mean of the distribution of individual natural frequencies. In the non-reciprocal case, the couplings learn to utilize the variance of the distribution of $\omega_a$ to achieve the desired global response:  oscillators with small or negative frequencies are strongly influenced by high positive frequency oscillators, but not vice versa, so that units far from the target frequency are pulled up without exerting a deleterious reciprocal effect. We can make this argument more precise by considering the mean frequency of all oscillators in system. Following the dynamical equation Eq.~\eqref{eq:Kuramoto_dyn}, we may write 
\begin{equation}
    \langle\dot{\theta}\rangle = \frac{1}{N} \sum_a \dot{\theta}_a =  \langle\omega\rangle - \frac{1}{N}\sum_{a,b}  K_{ab} \sin(\theta_a-\theta_b).
\end{equation}
For a reciprocal network, $K_{ab} = K_{ba}$, and the final term cancels pairwise, such that $\langle\dot{\theta}\rangle = \langle\omega\rangle$. 
Since $\omega_a$ are selected from a uniform distribution and held fixed during training, the mean value of $\dot{\theta}_a$ likewise \emph{cannot} change; so the task of synchronizing at $\omega_\text{sync} \neq \langle\omega\rangle \approx 0$ is inaccessible to reciprocal networks.

This example demonstrates the importance of non-reciprocal interactions: some kinds of tasks require them, and hence require extensions beyond the standard, static, energy-based physical learning paradigms.

\subsection{Leaky Integrate-And-Fire networks}
\label{sec:LIF}

Next, we consider a network of leaky integrate-and-fire (LIF) neurons connected by non-reciprocal synapses.  Networks of this kind are  used as coarse descriptions of neural circuits in the brain~\cite{dayan2005theoretical,teeter2018generalized} in situations where detailed Hodgkin-Huxley~\cite{hodgkin1952quantitative} models are not needed. Similar models are also used to build recurrent neural networks (RNNs) in software and have been realized in neuromorphic hardware~\cite{nahmias2013leaky,rozenberg2019ultra}. The dynamics of an LIF neuron $a$ with activation $x_a$ is given by
\begin{equation}
\begin{aligned}
&\dot{x}_a=\sum_b W_{ab} s(x_b,\beta_b) - \lambda_a x_a\\
&s(x_b,\beta_b)=[1+\textrm{exp}(-x_b-\beta_b)]^{-1}
\end{aligned}
 \label{eq:LIF}
\end{equation}
where $W_{ab}$ is a non-symmetric matrix of synaptic weights, $\beta_a$ is the bias of each neuron, $\lambda_a$ its leakiness, and $s$ is a sigmoid function describing the response nonlinearity. In our representation, the neuronal activations $x_a$ are the physical DOF, while the weights $W_{ab}$, biases $\beta_a$ and leaks $\lambda_a$ are three distinct sets of tunable DOF.

\begin{figure}
\includegraphics[width=1.0\linewidth]{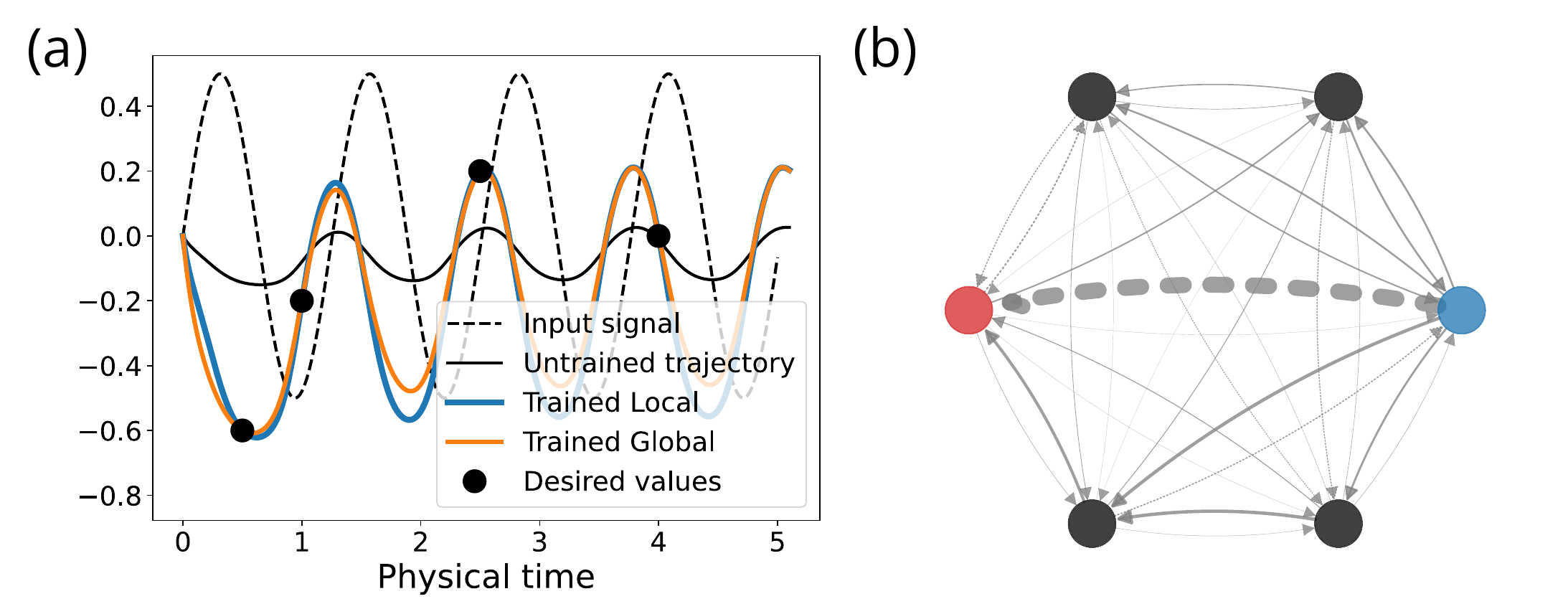}
\caption{Training a leaky integrate-and-fire neural network to reproduce points on a dynamical trajectory. ({\bf a}) A network of $N=6$ nodes with initially random non-reciprocal connections is trained such that when a periodic signal is applied to the input node (black dashed line), the output node exhibits particular values at specified times (black dots). The initial output before training (solid black) changes to reproduce the target trajectory after training with the forward supervisor (blue) as well as the gradient supervisor (orange). ({\bf b}) Network connectivity after training with the local rule (Eq.~\eqref{eq:LR3}) and the forward supervisor (Eq.~\eqref{eq:NewSup2}). The input neuron is red, the output neuron is blue, hidden neurons are black. The strength of a directed connection between two nodes is indicated by the edge width, with excitatory (positive) connections shown as solid lines and inhibitory (negative) connections shown as dashed lines.
\label{fig:figTrajectory}}
\end{figure}

First, we implement a dynamical {\it target  task}: apply a sinusoid  at an input neuron (dashed line in Fig.~\ref{fig:figTrajectory}a) and train a designated output neuron to take particular values at four specific times (black dots in Fig.~\ref{fig:figTrajectory}a).   We  start with $N=6$ fully connected LIF neurons with tunable DOF  drawn independently from a normal distribution, with $W_{ab}\sim\mathcal{N}(0,4)$, $\beta_{a}\sim\mathcal{N}(0,1)$, $\lambda_a\sim U(0,1)$ where $\mathcal{N}(\mu,\sigma^2)$ is a mean $\mu$ normal distribution with standard deviation $\sigma$, and $U(n,m)$ is uniform between $n$ and $m$.  The initial condition is  $x_a=0$.  Initially, the untrained output activation trajectory does not satisfy the task and is coupled weakly to the driving signal. We iteratively train the network with our local learning protocol and with global gradient descent. Both learning methods produce an output satisfying the task (blue and orange curves in Fig.~\ref{fig:figTrajectory}a). An important structural feature develops in the network as it learns: it effectively loses the direct connection from the input to the output node and learns a strong inhibitory connection going back from the output to the input; the network learns to use hidden node connectivity to perform the task (Fig.~\ref{fig:figTrajectory}b).

\begin{figure}
\includegraphics[width=1.0\linewidth]{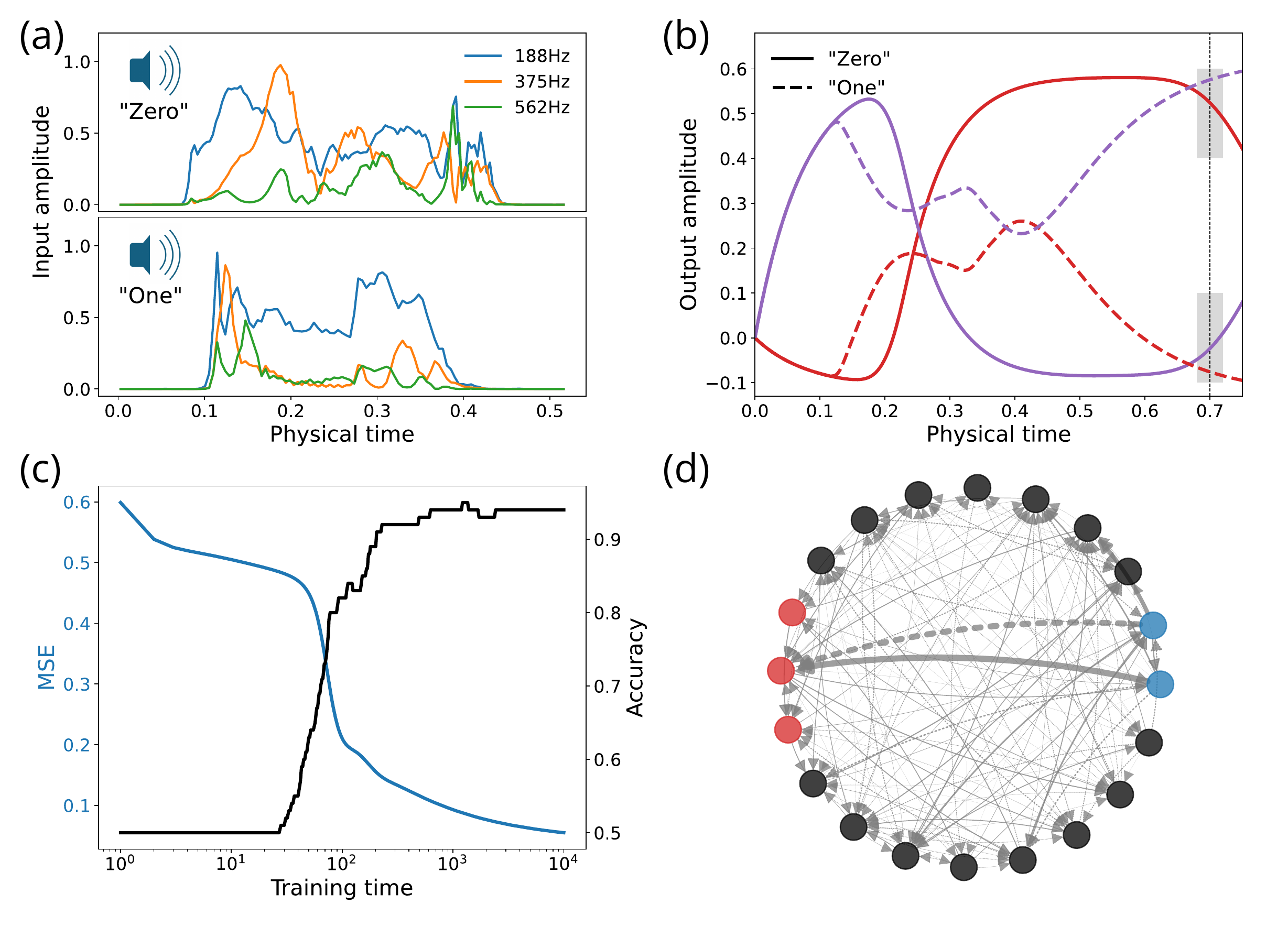}
\caption{Training a leaky integrate-and-fire neural network for audio MNIST classification with the forward supervisor. ({\bf a}) We provide three frequencies from spectrograms of a person saying  ``zero" and ``one" as time-dependent signals to three input nodes. The output amplitude is read out at two nodes. ({\bf b}) The network learns successfully  to exhibit high/low amplitudes at the Zero (solid line) and One (dashed lines) nodes at time $t=0.7s$ when the input is ``zero" (red lines), and low/high amplitudes, respectively, when the input is ``one" (purple lines). ({\bf c}) The dynamical learning process reduces the mean squared error on the test set (blue line), and increases the accuracy on the test set (black line)  $0.5$ to $0.95$. ({\bf d}) Network connectivity after training with the local rule (Eq.~\eqref{eq:LR3}) and the forward supervisor (Eq.~\eqref{eq:NewSup2}) (input neurons = red; output neurons = blue; hidden neurons = black).  Edge widths show the strength of directed connections between neurons, with excitatory (positive) connections shown as solid lines and inhibitory (negative) connections shown as dashed lines. Here, training creates strong directed connections between the two output neurons and the input neuron encoding the $375Hz$ spectrogram component, which well separates the `Zero' and `One' audio examples.
\label{fig:figAMNIST}
}
\end{figure}

We also train an LIF  network to perform {\it audio classification} of the downsampled Audio-MNIST dataset~\cite{audiomnist2023} of sound forms of multiple speakers pronouncing the  English digits $0-9$. As input we choose one  person (speaker $\#12$) and decompose their pronunciations `zero' and `one' ($50$ examples each) into  spectrograms. We chose $3$ frequencies from these spectrograms ($188, 375, 562$ Hz) and extract the associated audio time series (Fig.~\ref{fig:figAMNIST}a). We use data for two utterances of ``zero" and ``one" as training data, added as time-dependent driving signals to three arbitrary input neurons, one for each frequency, in an $N=20$ LIF neuron network. To define the task (Fig.~\ref{fig:figAMNIST}b), we choose two arbitrary output nodes, labeled ``Zero" and ``One." We train the ``Zero" node to have high activation and the `One" node to have  low activation at time $t=0.7$ if the input corresponds to the word ``zero." The roles of the ``Zero" and ``One" nodes are reversed if the input corresponds to ``one."  This is similar to one-hot encoding  used for classification tasks in machine learning. We randomly initialize our networks such that only half of the synaptic connections are present. The tunable DOF are initially drawn independently from distributions with $W_{ab}\sim\mathcal{N}(0,1)$, $\beta_{a}\sim\mathcal{N}(0,1)$, $\lambda_a\sim U(0,3)$. The initial condition for the dynamics is $x_a=0$ for all nodes.

Iterative training with our local learning rule produces the desired output response on the test set (Fig.~\ref{fig:figAMNIST}b) -- error on the training set drops, and  classification accuracy for the 48 utterances in the test set  improves from $50\%$ to $95\%$ (Fig.~\ref{fig:figAMNIST}c).   As above, learning produces an interesting structural feature in the trained network (Fig.~\ref{fig:figAMNIST}d): the network develops strong directional connections between the output nodes and one of the three input nodes, associated with the $375Hz$ spectrogram component (Fig.~\ref{fig:figAMNIST}a), which most strongly distinguishes the `Zero' and `One' examples. The network thus learns to couple the most useful input signal to the outputs to perform the classification task.

\begin{figure}
\includegraphics[width=1.0\linewidth]{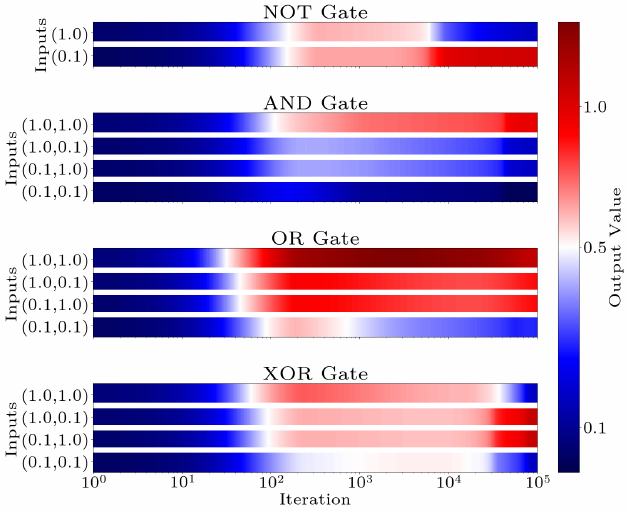}
\caption{Training a system of reacting chemical species to learn Boolean logic gates. A network of $N=10$ species, reacting following biochemical kinetic dynamics Eq.~\eqref{fig:MM}, is trained to reproduce unary NOT, and binary AND, OR, and XOR logic gates. A logical ``TRUE" for a given node (species) is indicated by a concentration $x=1.0$ (arbitrary units) while ``FALSE" is indicated by $x=0.1$.  Early in training, species do not react strongly due to initialization, with the output species always showing a logical ``FALSE" for all inputs (dark blue; left side of figure). After training, each logic gate settles at the correct Boolean output for the desired gate (red for ``TRUE," blue for ``FALSE"; right side of figure).
\label{fig:MM}
}
\end{figure}

\subsection{Biochemical reaction kinetics}
\label{sec:MM}
We next consider a tunable system of interacting biochemical species subject to enzyme-catalyzed Michaelis-Menten dynamics~\cite{michaelis1913kinetik}. Suppose we have $N$ species $\{X_a\}$, interacting via effective reactions of the form $X_a \xrightleftharpoons{X_c} X_b$, in which species $X_a$ and $X_b$ inter-convert. To allow for development of enzyme inhibition, we envision each reaction proceeds through a variety of channels, with each channel mediated by a third species, $X_c$, whose action is purely inhibitory. As a general model meant to capture the essence of such an interacting system, we consider the following dynamics,

\begin{multline}
    \dot{x}_a = \sum_{b\neq a} \sum_c \bigg( \frac{V_{ab}^c x_b}{R_{ab}^c+x_b+\alpha_{ab}^cx_c} -   \frac{V_{ba}^c x_a}{R_{ba}^c+x_a +\alpha_{ba}^cx_c} \bigg)\\
    - d x_a 
    \label{eq:MM_dynamics}
\end{multline}
where $x_a$ represents the concentration of species $X_a$, $V_{ab}^c$ is the limiting rate at substrate saturation for the $X_c$--mediated reaction channel $X_b \xrightarrow{X_c} X_a$, $R_{ab}^c$ is the corresponding Michaelis constant, $\alpha_{ab}^c$ parametrizes the inhibitory effect of $X_c$ on this reaction channel, and $d>0$ represents a fixed universal effective decay rate for all species. There are three sets of tunable degrees of freedom: $V_{ab}$, $R_{ab}$ and $\alpha_{ab}$ all subject to the constraint that they are positive. With these definitions, the first two terms of Eq.~\eqref{eq:MM_dynamics} represent, respectively, the production of species $X_a$ due to reactions of the form $X_b\xrightarrow{X_c} X_a$ and the consumption of $X_a$ due to reactions of the form $X_a\xrightarrow{X_c} X_b$. We initialize the tunable degrees of freedom randomly with $V_{ab}^c \sim \mathcal{N}(0,0.01),0, \, R_{ab}^c\sim\mathcal{N}(1,0.01),$ and $\alpha_{ab}^c \sim \mathcal{N}(1,0.01)$, ensuring that all $V_{ab}^c, R_{ab}^c, \alpha_{ab}^c$ are positive by setting any randomly selected negative initialization to zero. This initialization means that initially no chemical reactions are active and all species act inhibitively on each other. As a result, the fixed-point output concentration before tuning is small.

Taking inspiration from prior work~\cite{Erbas2018MolecularLogic}, we consider the task of training a fully-connected network of $N=10$ species to construct logic gates.  Unlike our previous examples, the dynamics in Eq.~\eqref{eq:MM_dynamics} involve a three-body interaction and thus the network topology for this system is a fully-connected three-body hypergraph: local updates will likewise be three-body in nature. For our task, we select a single chemostatted reference species fixed at $x_\text{ref} = 1$, a relevant number of input species to be used as logical inputs chemostatted either at logical ``TRUE," for which $x_\text{True}=1.0$, or logical ``FALSE," $x_\text{False}=0.1$, and one output species, whose fixed point after a settling time $T=5$ is used as a logical output. All remaining species are left as hidden.

In each iteration of training, we reset the hidden and output species concentrations to 0 at the beginning. In the free state we then follow the reaction kinetics over a duration of $T=5$ and observe the final concentration of a selected output species. In the clamped state we nudge the concentration of the output species over the entire trajectory towards the desired output concentration indicated by ``True" ($x_O=1.0$) or ``False" ($x_O=0.1$).  

In Fig.~\ref{fig:MM}, we plot the results for tuning the biochemical dynamics to express a number of different logic gates. In all cases, we use the local rule (Eq.~\eqref{eq:LR3}) and the forward supervisor (Eq.~\eqref{eq:NewSup2}) to tune $V_{ab}^c, R_{ab}^c$, and $\alpha_{ab}^c$ over 100,000 iterations of training. After sufficient training, the system is able to reproduce all logic gates. Note that the set $\{\text{NOT}, \text{AND} \}$ form a functionally complete logical set \cite{Enderton1972Logic}; so, in principle, any logical function can be expressed in this system after training.

\begin{figure*}
\includegraphics[width=1.0\linewidth]{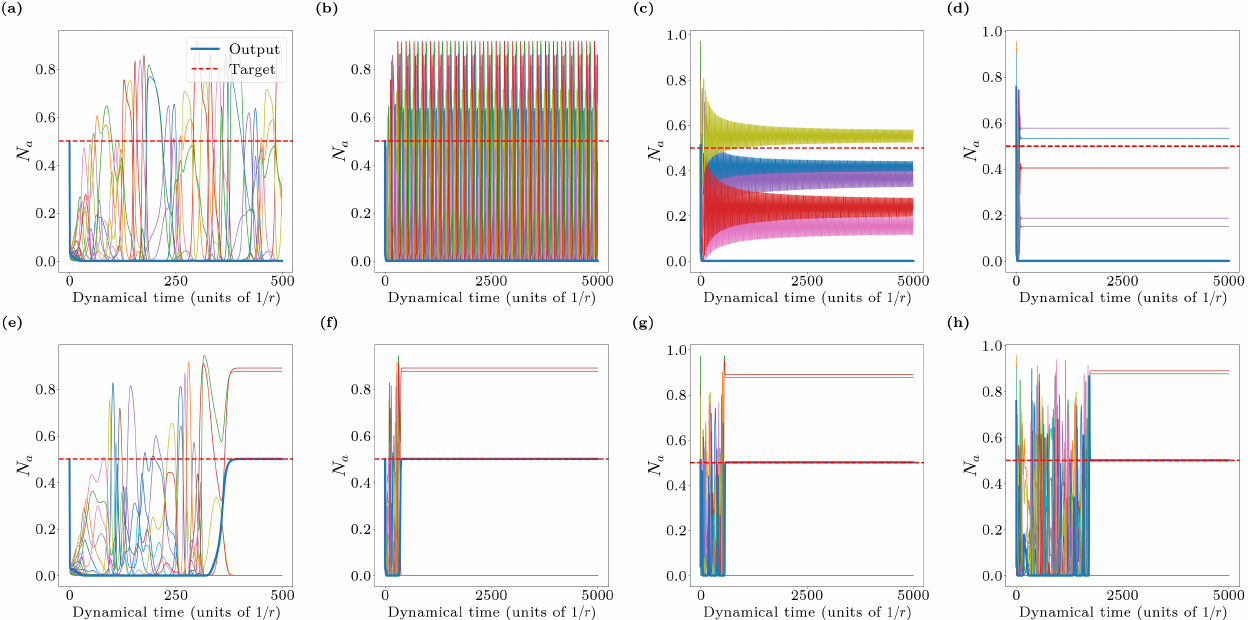}
\caption{Training a system of interacting ecological species to stably fix at a desired target. We study a fully-connected network of $S=50$ highly fluctuating species following generalized ecological dynamics Eq.~\eqref{eq:LV_dynamics}. We choose intrinsic carrying capacities of $K_a=1$ and intrinsic growth rates $r_a=1$ for all species, such that we describe normalized species counts $N_a$ and measure dynamical time in units of $1/r$. We then desire one selected species (highlighted in dark blue in all figure panels) to fix at a target normalized species count of $N_{\text{target}} = 0.5$ (shown in dashed red). We train iteratively over a duration $T=500$ and for initial conditions $N_0 = 0.5$ for all species. In each column of the figure, the systems have the same initial conditions. 
(\textbf{Top Row}) Before training: 
(\textbf{a}) The network is fluctuating strongly over the training duration $T=500$ for the selected training initial conditions.
(\textbf{b}) For the same initial conditions, but over a longer time horizon, the system exhibits a quasiperiodic limit cycle. 
(\textbf{c}) \& (\textbf{d}) For different choices of initial conditions, the system settles into other attractors, such as periodic limit cycles or fixed points.
(\textbf{Bottom Row}) After training: 
(\textbf{e}) After iterative training, the network designs a new fixed point attractor that it reaches within the training duration of $T=500$.
(\textbf{f}) For the same initial conditions as (e), the fixed point extends stably beyond the training window.
(\textbf{g}) \& (\textbf{h}) For the same initial conditions as in (c) \& (d), respectively, the network generalizes, settling into the same fixed point with the output species on-target.
}
\label{fig:LV}

\end{figure*}

\subsection{Ecological dynamics}
\label{sec:LV}

As a final example, we consider tunable models of the ecological dynamics of a large number of interacting, competing species \cite{Lotka_Volterra, Bunin2017LV}. Taking inspiration from a recent work on competition in bacterial colonies \cite{Gore2022LV}, we consider the following dynamics 
\begin{equation}
  \dot{N}_a =  \frac{r_a}{K_a} N_a \left(K_a - N_a - \sum_{b \neq a} ^SA_{ab} N_b \right) + D_a
    \label{eq:LV_dynamics}
\end{equation}
where $S$ is the size of the species pool, $N_a$ is the population of species $a$, $K_a$ is the corresponding carrying capacity in the absence of interactions, $r_a$ is a species-specific growth rate, $D_a$ is an invasion rate, and $A_{ab}$ is an interaction matrix, with positive values indicating predation of species $b$ on species $a$. For illustration we focus on a case for which $r_a=K_a=1$ and $D_a = 10^{-6}$ are fixed for all species, such that $N_a$ represents the \textit{normalized} species count or abundance of species $a$. We allow $A_{ab}$ to be the tunable degrees of freedom. In the case where the $A_{ab}$ are drawn from a uniform distribution, $U(0,2\mu)$, \cite{Gore2022LV} demonstrated that these dynamics display three primary physical phases determined by $\mu \equiv \langle A_{ab} \rangle$: (I) for sufficiently small $\mu$, a phase with a unique fixed point in which all species co-exist, (II) for intermediate $\mu$, a phase with a unique fixed point in which only a fraction of species co-exist, and (III) for sufficiently large $\mu$ a phase with multiple dynamical attractors, including multiple possible fixed points or persistent fluctuating attractors set by initial conditions (see Appendix~\ref{app:eco}). 

As a task meant to both be relevant to these system dynamics and to challenge the limits of the learning paradigm, we train a system initially in Phase (III) (with multiple attractors) so that a specific output species fixes at a desired value $N_\text{output}^d=0.5$. We consider a pool of $S=50$ species and initialize $A_{ab} \sim U(0,2)$, and select a dynamical species initialization $N_a(0)=0.5$. These choices ensure the system is initialized in Phase (III) \cite{Bunin2017LV,Gore2022LV}. As an output, we select a species that is not active for these selected initial conditions before tuning. We then train the system by tuning the interaction matrix $A_{ab}$ iteratively over 10,000 iterations, each of duration $T=500$. At the onset of each iteration, we re-initialize all the physical degrees of freedom to a fixed abundance of $N_{0}=0.5$ for all species.

In Fig.~\ref{fig:LV}a, we plot the system over $T=500$ time units for these initial conditions, $N_a(0) = 0.5$, before training. In Fig.~\ref{fig:LV}b, we plot the system with the same initial condition over a much longer time horizon, showing that these conditions lead to a limit cycle attractor. In Fig.~\ref{fig:LV}c and d we plot the system for two different initial conditions drawn from a uniform distribution, $N_{a} \sim U(0,1)$. Each of these different initial conditions leads to a different attractor. Note that in each of these attractors, the chosen target species, shown in thick blue, does not actively participate. The presence of multiple attractors confirms that the system is indeed in Phase (III) (see Appendix~\ref{app:eco} for a full catalog of all attractors). 

After sampling 10,000 random initial conditions, we find that the chosen target species \text{does} appear alive 1041 times for these dynamics before tuning, always due to a specific fixed point attractor with $N_\text{output}^* = 0.67$. We set the desired value of the target species to $N_\text{output}^d=0.5$ in order to avoid this attractor.

Fig.~\ref{fig:LV}e-h replicate Fig.~\ref{fig:LV}a-d, with the same initial abundances used in each column of Fig.~\ref{fig:LV}, but for the system after training. Specifically, Fig.~\ref{fig:LV}e and f demonstrate that the target species settles at the desired abundance of $N_\text{output}^d=0.5$, even well beyond the training time window of $T=500$ (Fig.~\ref{fig:LV}f). Further, as shown in Figs.~\ref{fig:LV}g and h, although training was only carried out for the specific initial condition of $N_0=0.5$ for all species, the results generalize to other initial conditions--for each of these, $N_\text{output}^d=0.5$. Checking the same randomly selected 10,000 initial conditions, the target species now fixes to the desired value in 8333 out of 10,000 cases. The existence of other attractors show that the system remains in Phase (III) even after training (see Appendix~\ref{app:eco}). 

Taken together, training in this highly fluctuating and dynamical case amounts to shaping the dynamical attractor basins of this system. Previous work \cite{Guzman2024Imprints,stern2025physical} for the static and reciprocal energy-based systems has investigated the effects of tuning on the Lyapunov function landscape for fixed point tasks in systems with reciprocal interactions; this example raises the corresponding question of how tuning informs the dynamical attractor landscape more generally. We leave this exploration to future study.

\section{Gradient Alignment}

\begin{figure*}
    \centering
    \includegraphics[width=1.0\linewidth]{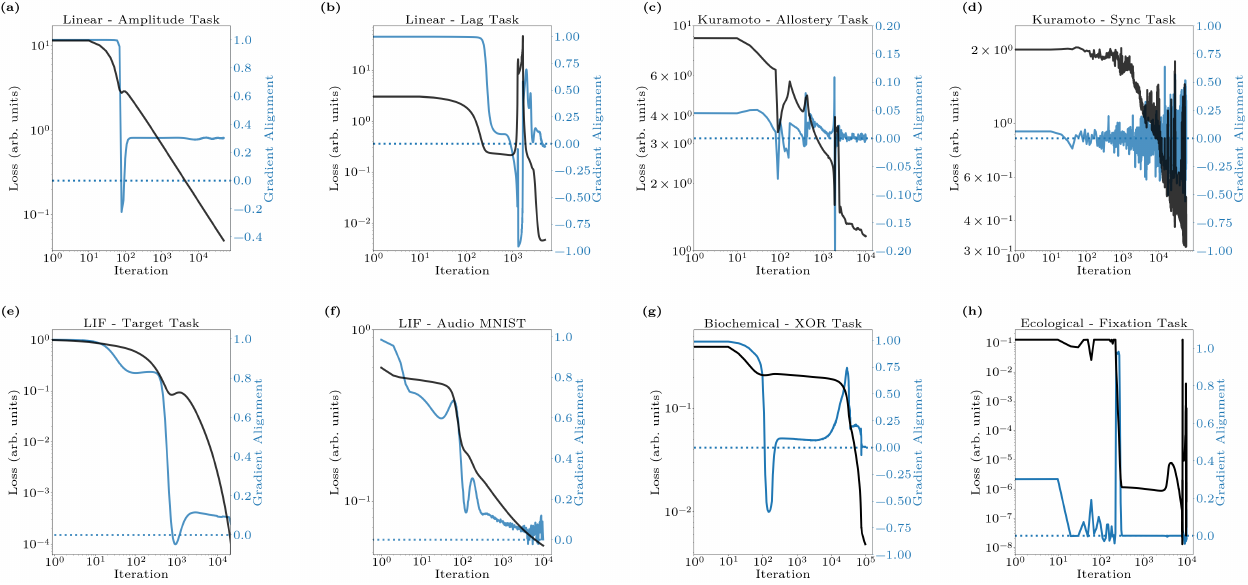}
    \caption{{\bf Black curves}: The cost functions of executing each of the target tasks described in Sec.~\ref{sec:demonstrations} decrease as training progresses using the local rule and forward supervisor.  {\bf Blue curves}: the alignment (the normalized dot product) between the cost gradient and the weight updates of Eq.~\ref{eq:LR3} with the forward supervisor of Eq.~\ref{eq:NewSup2}. Training is successful with the local forward process despite alignment fluctuations and complex dynamics. The horizontal dashed line corresponds to vanishing alignment.    
    ({\bf a})  {\it Amplitude task} for  linear oscillators  with symmetric couplings.  We train the time-dependent displacement of an output node to mirror the displacement of an input node, amplified by a factor $p$ (Sec.~\ref{sec:CoupledLinear}).  
    ({\bf b}) {\it Lag task} for linear oscillators.  We train the displacement of an output node to mirror an applied stimulus at an input node, with a time lag $\Delta t$ (Sec.~\ref{sec:CoupledLinear}).    
    ({\bf c}) {\it Allostery task} for a Kuramoto oscillator network.  We train the network to synchronize the phase of an output node with the time-dependent phase of an input node (Sec.~\ref{sec:Kuramoto}).
    ({\bf d})  {\it Synchronization task} for a Kuramoto oscillator network.  We train the network to synchronize globally at a frequency $\omega_{\text{sync}} = 1$, far above the mean of the distribution of intrinsic oscillator frequencies (Sec.~\ref{sec:Kuramoto}). 
    ({\bf e}) {\it Target dynamical task} for a Leaky Integrate-and-Fire neuron network.  We train an output neuron to reproduce points on a target trajectory driven by a sinusoidal signal at an input neuron  (Sec.~\ref{sec:LIF}).
    ({\bf f}) {\it Audio classification task} for a Leaky Integrate-and-Fire neuron network.  We train the network to classify whether the input (the time series of three frequencies in utterances from the Audio-MNIST database) is drawn from a spoken ``one" or ``zero," by producing a large response in one of two output neurons at a specified time (Sec.~\ref{sec:LIF}).
    ({\bf g}) {\it XOR task} for a biochemical network. We train the network to replicate the Boolean logic function exclusive-or as a function of specified input species concentrations (Sec.~\ref{sec:MM}).
    ({\bf h}) {\it Fixation task} for an ecological network. We train an output species to stably fix at a specified value in a multi-stable parameter regime with multiple attractors (Sec.~\ref{sec:LV}).
    }
    \label{fig:Alignment}
\end{figure*}

As we discussed in Sec. II, the local rule, Eq.~\eqref{eq:LR3}, with the forward supervisor, Eq.~\eqref{eq:NewSup2}, does not reproduce  gradient descent. Nonetheless, we showed in Sec. III that training is successful for various  systems and tasks. This suggests that our weaker learning condition, that the local learning process correlates with the gradient on average, holds for our protocol, even for cases far from the regime that we can understand analytically. To verify for the examples studied in Sec~\ref{sec:demonstrations} we calculated: (1) the cost function (loss) during training (black lines in Fig.~\ref{fig:Alignment}), and (2) the alignment (normalized dot product) between the modification resulting from our local learning protocol and the global cost gradient (blue lines in Fig.~\ref{fig:Alignment}). 

In Fig.~\ref{fig:Alignment} we observe the cost decreasing over training time, sometimes by orders of magnitude, as  task performance increases significantly. However, the cost reduction is not monotonic over training time, with sporadic bouts of increasing cost. Such cost increases could arise for two reasons: (1) the learning rate is too high, so that the tunable DOFs change too much in some training steps (a source of such non-monotonicity in training error even for gradient descent), or, (2) because of misalignments between the cost gradient and the changes produced by our local learning protocol. 

Fig.~\ref{fig:Alignment} illuminates the relation between aligment and cost reduction.   For the linear (Fig.~\ref{fig:Alignment}a,b), LIF (Fig.~\ref{fig:Alignment}e,f), and biochemical (Fig.~\ref{fig:Alignment}g) networks, alignment between the cost gradient and the local update direction is typically positive during training, and the cost decreases consistently. When the alignment occasionally becomes negative (implying that our local protocol is anti-aligned with the cost gradient), there is some increase in cost, as expected. Interestingly, the bouts of anti-alignment in these examples are short in duration, and local learning quickly realigns with the gradient, allowing the learning process to recover. Such complex alignment dynamics could have many sources; we will leave a systematic study to future work. That said, these linear, LIF, and biochemical networks illustrate the core intuition of our PAR protocol: learning is successful because the local learning protocol typically correlates with the global cost gradient.

The Kuramoto oscillator (Fig.~\ref{fig:Alignment}c, d) and ecological (Fig.~\ref{fig:Alignment}h) networks present examples where the local learning protocol correlates more poorly with the global cost gradient. In the Kuramoto allostery task (Fig.~\ref{fig:Alignment}c), the alignment is initially mildly positive, meaning that the local rule is weakly aligned with the cost gradient. However, as training progresses, the alignment fluctuates between small positive and small negative values, accompanied by associated fluctuations in cost. Although the local rule never strongly aligns with the gradient, the loss steadily decreases over training. The Kuramoto synchronization task (Fig.~\ref{fig:Alignment}d) likewise starts with a period of weak positive alignment during which the cost is roughly constant. But then the alignment begins to fluctuate rapidly between positive and negative values, while  the cost decreases rapidly with occasional sharp fluctuations. A similar dynamic occurs in the ecological network (Fig.~\ref{fig:Alignment}h): the alignment is generally positive, but fluctuates substantially, leading to a cost that ultimately decreases with notable momentary sharp increases. In all of these cases, learning is successful because, on average, there is a bias towards positive alignment. These examples demonstrate cases where our weak PAR condition requiring  positive alignment with the gradient \emph{on average} is  sufficient for learning. We leave further study of the local learning dynamics and alignment with the global cost gradient to future work.

\section{Discussion}

We have introduced a physical learning paradigm that extends earlier approaches based on local learning rules for equilibrium or steady state tasks in systems with reciprocal interactions. Our paradigm uses spatially local updates to train dynamical systems governed by coupled ordinary differential equations, relying on physics to propagate information causally.  Our protocol involves a local learning rule and a causal supervisor acting only on output nodes. The new rule (Eq.~\eqref{eq:LR3}) and supervisor (Eq.~\eqref{eq:NewSup2}) reduce to standard contrastive learning when applied to reciprocal networks trained for static tasks.

In the slow learning rate limit, our local rule can recover exact gradient descent on a cost function with appropriate supervision. However, the required supervision is not tractable for large system sizes -- the supervisor must nudge all the physical degrees of freedom, not just the output nodes, at all times towards the desired trajectory. This is because errors at each time are propagated to all future times by the physical dynamics. Thus, to correct  errors at a given time, the supervisor has to backpropagate adjustments to every node at every earlier instant.

To define a \emph{tractable} supervision protocol, we note that a successful training process does not always have to follow the gradient perfectly.  It just has to get it approximately right often enough.  This leads us to Probably Approximately Right (PAR) supervision.  A weak form of PAR supervision requires positive correlation on average between learning induced updates and the global cost gradient.  We propose a tractable supervisor coupled with a physically local learning rule, and demonstrate analytically that it satisfies our weak PAR condition for some simple tasks.

We have not determined general conditions under which our proposed local learning process for dynamical systems satisfies the PAR condition and leads to successful learning.  It would be interesting to derive such results.  Here, we provided evidence for the applicability of our protocol by demonstrating it for several classes of tunable dynamical networks: coupled linear oscillators, Kuramoto networks, networks of Integrate-And-Fire (LIF) neurons, Michaelis-Menten biochemical reaction networks and generalized Lotka-Volterra models.  We showed numerically that the PAR condition is satisfied for each of these systems and tasks.
 
Overall, our protocol aligns with the global cost gradient, but shows complex training dynamics.  Understanding the latter will be interesting for the future.

It is possible that the PAR supervisor works for our systems because they are sufficiently small and the tasks we are training them to perform are simple; it could be that as task complexity and system size increase, training processes must approach gradient descent more closely for successful learning. However, a different scenario is possible. If the system is sufficiently overparameterized, a PAR process might allow it to skate around on a slightly bumpy and low-lying region of the cost landscape to reach acceptable performance.  Note that the cost function itself cannot be defined uniquely; there could be many definitions of cost with the same solutions (zeros). Moreover, in biological systems, it is difficult to know the cost. If the cost function is non-unique or approximate, its gradient will be also. This suggests that a focus on processes that reduce to gradient descent in some limit might be less productive than a focus on PAR processes. This is particularly important for systems that implement learning and computation using physical processes as they are inevitably imperfect and cannot achieve digital precision. 

Our approach to physical learning has several applications. First, we can derive learning rules and supervisors for engineered systems~\cite{du2025metamaterials} learning dynamical trajectories autonomously while adapting to changing conditions and tasks. Such abilities will be useful, for example, for micro-robotic systems deployed in changing environments without on-board computers to reprogram behavior. Second, our approach could be embedded in adaptive mechanical systems for medical applications, implementing mechanical force generation~\cite{arzash2025rigidity,banerjee2025learning} (extracellular matrix, muscle) or required functions in pulsatile flow networks~\cite{chatterjee2025hierarchical,altman2025collective} (arteries, heart). Finally, our approach can be implemented in electrical circuits, enabling learning of tasks breaking time-reversal symmetry in addition to static ones~\cite{dillavou2022demonstration,dillavou2024machine}. In particular, our learning rule and supervisor should be useful for harnessing neuromorphic systems implementing leaky integrate-and-fire models with subthreshold transistors that have been difficult to train in a supervised manner~\cite{christensen20222022}.

Dynamical learning machines may also offer a new window into functional adaptation in living systems~\cite{stern2024physical}.  For example, circuits in the brain display structures conserved between individuals and species that are evolved or learned adaptations to the environment or to tasks, for example, in circuits supporting vision \cite{ratliff2010retina,garrigan2010design,hermundstad2014variance}, audition \cite{lewicki2002efficient,smith2006efficient}, olfaction \cite{tecsileanu2019adaptation,krishnamurthy2022disorder}, and spatial cognition \cite{wei2015principle}.   Some living systems  adjust by comparing different system states (e.g., brain~\cite{bengio2015stdp}, slime molds~\cite{marbach2023vein}, molecular reaction networks~\cite{trifonova2025trainable} and intracellular phase separated organelles~\cite{shin2017liquid}). In a paradigmatic example, the juvenile songbird learns the parental song by comparing it with its own developing babble, and efficient learning requires the ``teacher circuit'' in the brain to adapt its messages to the local plasticity rule implemented by the ``student circuit'' \cite{tecsileanu2017rules}. In such cases, our framework might lead to interpretation of biologically-plausible adaptation rules in terms of contrastive learning rules paired with supervisors, possibly enabling us to disentangle the local rule from the supervisor. Even in living systems that do not use contrastive learning, such as developing embryos undergoing morphogenesis, our protocol can be used to obtain ensembles of systems tuned for the same behavior, which can then be studied to gain insight into what the system is trying to learn~\cite{Arzash2025Epithelial} and how it is able to do so~\cite{stern2025physical,Guzman2024Imprints}. The ability to tune biologically-inspired models is a potentially valuable tool in our arsenal for understanding the link between microscopic properties and complex collective functionality in living matter.

\subsection*{Acknowledgments}
\noindent We thank Sam Dillavou, Cyrill Bosch, Vidyesh Rao Anisetti, Yao Du and Corentin Coulais for useful discussions. This research was supported by DOE Basic Energy Sciences through grant DE-SC0020963 (MS), NSF-DMR-MT-2005749 (AF, AJL) and the UPenn MRSEC DMR-2309043 (AF).  VB and MS were  supported by NSF grant CISE 2212519.  The Simons Foundation provided support via Investigator Award \# 327939 (AJL). 

\appendix

\section{Explicit dependence of the cost on the tunable degrees of freedom}
\label{sec:ExplicitDep}

In the main text we focused on learning dynamical behaviors that depend on the trajectories of the physical degrees of freedom $x^F_a(t,\vec w)$, and therefore only implicitly on the tunable degrees of freedom (DOF) $w_i$. We did not consider cost functions with explicit dependence on the tunable DOF $C(\vec x^F, \vec w)$. Such cost functions are common, and typically appear when one wishes to encourage the tunable DOF to reach specific kinds of solutions by adding regularization~\cite{mehta2019high}. Common regularization schemes include norm regularization, e.g. $C=C_0 + \lambda  \sum_i w_i^\ell$, where for $\lambda>0$, the learning process penalizes the $\ell$-norm of the tunable DOF vector. In a physical context, one could define such explicit tunable DOF dependence to encourage the system to find energy efficient solutions~\cite{stern2024training}.

Here we generalize our contrastive local learning approach to include such explicit dependence on tunable DOF, with the following restriction: the cost function can only explicitly depend on the tunable DOF in a local fashion~\cite{stern2025physical}. In other words, the partial derivative of the instantaneous cost function at time $t$ for a particular tunable DOF $w_i$, $\frac{\partial c(t)}{\partial w_i} = f(w_i,x_{a\in\Gamma(i)}(t))$, can only depend on the value of that tunable DOF and the physical DOF in its local neighborhood $\Gamma(i)$--the physical DOF at the nodes connected by that tunable DOF--at that particular time. The regularization schemes discussed above, as well as many others, satisfy this condition. 

We start by computing the gradient of the total cost $C$ as in Eq.~\eqref{eq:CG1}, but now include explicit $\vec w$ dependence:

\begin{equation}
\begin{aligned}
\frac{dC}{dw_i}&=\int_{0}^{T} dt \frac{\partial c(\vec{x}^F)}{\partial x_a^F} \frac{d x^F_a(t)}{d w_i} + \int_{0}^{T} dt \frac{\partial c(\vec w,t)}{\partial w_i}
\end{aligned}
 \label{eq:CGA1}
\end{equation}

In the main text we showed how to deal with the first term in this gradient. The new second term is easier to account for: since the explicit dependence on the tunable DOF satisfies the locality condition above, one can redefine the local learning rule to include this term of the gradient explicitly in the tunable DOF dynamics of Eq.~\eqref{eq:LR3}:
\begin{equation}
\begin{aligned}
\Delta w_i &= \frac{\alpha}{\eta} \int_0^T  dt'  \, (\vec{x}^C(t') - \vec{x}^F(t'))
\cdot \frac{\partial \vec{F}(t')}{\partial w_i}\\
&-\alpha \int_{0}^{T} dt \frac{\partial c(\vec w,t)}{\partial w_i}
\end{aligned}
 \label{eq:LRA6}
\end{equation}
As an example, consider training a dynamical system for some properties of its trajectory, while also performing $L2$-regularization to penalize high norms for its tunable DOF vector $\vec w$. The cost function we wish to minimize is $C(x^F_a,w_i)=C_0(x_a^F)+\frac{1}{2}\lambda \sum_i w_i^2$, where $C_0(x_a^F)$ is the original cost function that depends on the dynamical trajectory but not explicitly on the tunable DOF. For this system, the appropriate learning rule is therefore $$\Delta w_i = \frac{\alpha}{\eta} \int_0^T  dt'  \, (\vec{x}^C - \vec{x}^F)
\cdot \frac{\partial \vec{F}}{\partial w_i} - \alpha \lambda T w_i.$$
The first term in this learning process is the one discussed in the main text. The second term comes from the minimization of the explicit cost dependence on the tunable DOF and is manifestly local in space and time. Therefore, this learning rule can potentially be encoded in the physics of a dynamical learning system.

\section{Correlations of local and global learning processes}

To evaluate the fidelity of a local learning rule, we calculate its projection onto the exact cost gradient, the \emph{normalized dot product} (cosine similarity) between their respective tunable DOF updates. Consider a dynamical system $\dot{\vec{x}} = \vec{F}(\vec{x}, \vec{w}, t)$ and a total cost $C = \int_{0}^{T} c(t) dt$. We will show that, under certain assumptions, the alignment between both updates approaches one, so that the local learning process well-approximates the gradient. The updates for the tunable DOF due to the local rule~\eqref{eq:NewSup2}, $\Delta w^L_i$, and due to gradient descent~\eqref{eq:CG2}, $\Delta w^G_i$, result from the interaction between parameter sensitivity $\vec \nabla_w \vec{F}$ and the propagated error signal $\vec{\nabla}_x c(t)$ weighted by the signal matrix $S$ computed in \eqref{eq:NudgePropagator}: 

\begin{equation} 
\begin{aligned}
    &\Delta w^L_i \propto \int_{0}^{T}\int_{0}^{T} dt \, dt' \, \frac{\partial F_a(t)}{\partial w_i} S_{ab}(t,t') \frac{\partial c(t')}{\partial x_b}   \\
    &\Delta w^G_i \propto \int_{0}^{T}\int_{0}^{T} dt \, dt' \,  \frac{\partial c(t)}{\partial x_a}  S_{ab}(t,t') \frac{\partial F_b(t')}{\partial w_i}
\end{aligned} 
\label{eq:wupdatedefs}
\end{equation}

To compare these vectors, we assume a discrete allostery task where the error is a pulse at $t^*$, i.e., $\frac{\partial c(t)}{\partial x_a} = e_a(x) \delta(t-t^*)$ for an error signal $e_a(x)$. We further assume the signal matrix $S$ decays rapidly, such that perturbations to node values only affect the dynamics for a short time beyond $t^*$, that we will denote by $\delta t$ (e.g. system dynamics rapidly decay to a fixed point). 
We may then expand $\Delta w^L_i$ for a small forward time step $\alpha = t - t^* > 0$. 

For $\Delta w^G_i$ we assume a rapid backwards decay of the temporal signal, expanding for a small backward step $\beta = t^* - t' > 0$. Since all times from the beginning of the trajectory to the time of the error signal can contribute to the error, this assumption only holds if the error is measured near the start of the trajectory, so that $t^*$ itsel is very small. In that case, the backwards signal is quickly cropped by the initial condition.\\

Under these assumptions, we arrive at the following expansions:
\begin{equation}
\begin{aligned}
&\begin{split}
S_{ab}(t^*+\alpha, t^*) \approx \delta_{ab} &+ \alpha \frac{\partial F_a(t^*)}{\partial x_b}   \\ 
&+ \frac{1}{2}\alpha^2\left[\frac{\partial \dot{F}_a(t^*)}{\partial x_b} 
+ \left(\frac{\partial F(t^*)}{\partial x} \right)^2_{ab}\right]
\end{split}\\
&\begin{split}
S_{ab}(t^*, t^*-\beta) \approx \delta_{ab} &  +\beta \frac{\partial F_a(t^*)}{\partial x_b}   \\ 
&+ \frac{1}{2}\beta^2\left[-\frac{\partial \dot{F}_a(t^*)}{\partial x_b} 
+ \left(\frac{\partial F(t^*)}{\partial x} \right)^2_{ab}\right]
\end{split}
\end{aligned} 
\label{eq:Sexpn}
\end{equation}
Similarly, we expand the $\frac{\partial F}{\partial w}$ around the error time $t^*$:
\begin{equation} 
\begin{aligned}
    \frac{\partial F_a(t^*+\alpha)}{\partial w_i} &\approx \frac{\partial F_a(t^*)}{\partial w_i} + \alpha \frac{\partial \dot{F}_a(t^*)}{\partial w_i} +\frac{1}{2} \alpha^2 \frac{\partial \ddot{F}_a(t^*)}{\partial w_i}
\end{aligned} 
\label{eq:fwexpn}
\end{equation}

Substituting Eqs.~\ref{eq:Sexpn}-\ref{eq:fwexpn} into Eq.~\eqref{eq:wupdatedefs} and integrating, keeping terms to second order in $\delta t$ and $t^*$, we obtain:
\begin{equation} 
\begin{aligned}
    &\frac{1}{\delta t}\Delta w^L_i \approx  v_i + \frac{1}{2} \delta t \, z_i + \frac{1}{6}\delta t^2 \,p_i \\
    &\frac{1}{t^* }\Delta w^G_i \approx  v_i + \frac{1}{2} (t^*) \,y_i +\frac{1}{6}(t^*)^2 \, q_i
\end{aligned} 
\label{eq:B4}
\end{equation}
where 
\begin{equation}\begin{aligned}
v_i &= \sum_a \frac{\partial F_a(t^*)}{\partial w_i} e_a(x)\\
z_i &= \sum_{ab} \left(\frac{\partial F_a(t^*)}{\partial w_i} \frac{\partial F_a(t^*)}{x_b} +\frac{\partial \dot{F}_b(t^*)}{\partial w_i} \right)e_b(x)\\ 
y_i &= \sum_{ab} e_a(x) \left( \frac{ \partial F_a(t^*)}{\partial x_b} \frac{\partial F_b(t^*)}{\partial w_i} - \frac{\partial \dot{F}_a(t^*)}{\partial w_i} \right) \\
\end{aligned},
\label{eq:B5}\end{equation}
where again $e_a(x)$ is the error signal, $\frac{\partial c}{\partial x_a} = e_a(x)\delta(t-t^*)$. 
Here, $p_i$ and $q_i$  are second order contributions which we will neglect in the following expansion.\\

The alignment, measured by the cosine similarity between updates, is obtained from:
\begin{equation} 
\begin{aligned}
\mathcal{A} \left(\Delta \vec w^L,\Delta \vec w^G\right)=\frac{\Delta \vec w^L\cdot \Delta \vec w^G}{\sqrt{\|\Delta 
\vec w^L\|^2\|\Delta \vec w^G\|^2}}
\label{eq:B6}
\end{aligned}
\end{equation}
Using our previous results (Eqs.~\ref{eq:B4}-\ref{eq:B5}) we find: 
\begin{equation}
    \begin{aligned}
        &\frac{1}{\delta t}\Delta \vec w^L\cdot \frac{1}{t^*}\Delta \vec w^G\approx \|\vec{v}\|^2 +\\
        & \frac{1}{2}\left(\delta t\vec{v}^T\vec{z}+t^*\vec{v}^T\vec{y}\right)+\frac{1}{6}\left(\delta t^2\vec{v}^T\vec{p}+t^{*2}\vec{v}^T\vec{q}\right)+\frac{1}{4}t^*\delta t\vec{y}^T\vec{z} \, \\
        & \delta t\|\Delta\vec w^L\|^{-1}\approx\left[\|\vec{v}\|^2+\delta  t \, \vec{v}^T\vec{z}+\delta t^2\left(\frac{1}{4}\vec{z}^T\vec{z}+\frac{1}{3}\vec{v}^T\vec{p}\right)\right]^{-\frac{1}{2}}
        \\
        &\approx\|\vec{v}\|^2-\frac{\delta t}{2}\vec{v}^T\vec{z}+\delta t^2\left[\frac{3}{8}\left(\vec{v}^T\vec{z}\right)-\frac{1}{8}\vec{z}^T\vec{z}-\frac{1}{6}\vec{v}^T\vec{p}\right],\\
        & t^{*} \|\Delta\vec w^G\|^{-1}\approx\left[\|\vec{v}\|^2+t^{*}\vec{v}^T\vec{y}+\left(t^{*}\right)^2\left(\frac{1}{4}\vec{y}^T\vec{y}+\frac{1}{3}\vec{v}^T\vec{q}\right)\right]^{-\frac{1}{2}} \\&\approx\|\vec{v}\|^2-\frac{t^{*}}{2}\vec{v}^T\vec{y}+\delta t^2\left[\frac{3}{8}\left(\vec{v}^T\vec{y}\right)-\frac{1}{8}\vec{y}^T\vec{y}-\frac{1}{6}\vec{v}^T\vec{q}\right],
    \end{aligned}
    \label{eq:B7}
\end{equation}
 Substituting Eq.~\eqref{eq:B7} into Eq.~\eqref{eq:B6}, we obtain the alignment to second order in $\delta t$ and $t^*$:
\begin{equation}
    \begin{aligned}
    &\mathcal{A} \left(\Delta \vec w^L,\Delta \vec w^G \right)\approx\\&\approx - \frac{1}{8 \|\vec{v}\|^4} \left\{ \|\vec{v}\|^2 \|\delta t\vec{z}-t^*\vec{y}\|^2-\left[\vec{v}^T\left(\delta t\vec{z}-t^*\vec{y}\right)\right]^2\right\}
        \end{aligned}
    \label{eq:B8}
\end{equation}

\begin{figure}[h!]
\includegraphics[width=1.\linewidth]{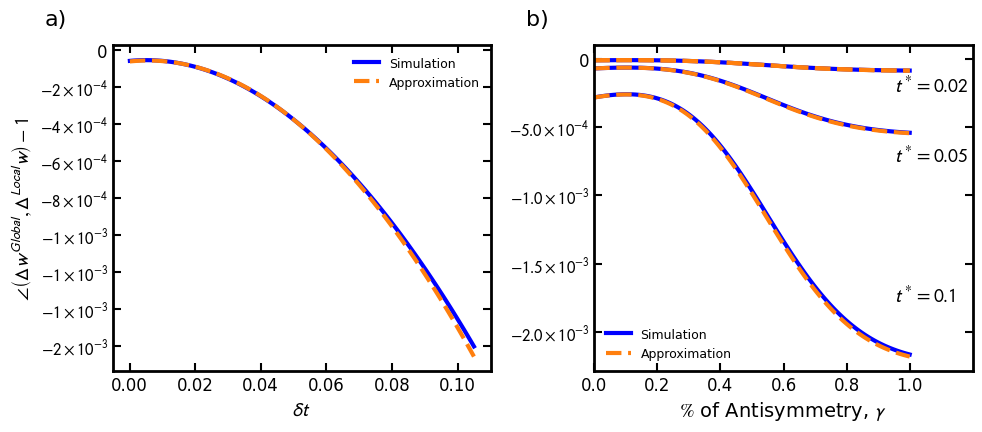}
\caption{Comparison of the misalignment between global gradient and local approximation $\mathcal{A}(\Delta^{\text{Local}} w, \Delta^{\text{Gradient}})$, based on our approximation (Eq.~\eqref{eq:B6}) and simulated data. a) Comparison for LIF dynamics (Eq.~\eqref{eq:LIF}) with a fixed error time $t^*=0.05$. Our analytical results well-approximate the simulated date to $\delta t\sim 0.1$. b) Comparison for linear dynamics (Eq.~\eqref{eq:linear_dynamics}), optimally setting $\delta t=t^*$. Pairs of curves for different values of $t^*$ with respect the symmetry of the dynamical matrix $A_\gamma$.}
\label{fig:CorrelationB}
\end{figure}

As reported in Section IID of the main text, the alignment between the cost gradient and its approximation tends to one under our assumptions ($t^*\ll 1$ and $\delta t \ll 1$), meaning that the local learning rule and tractable supervisor perform nearly exact gradient descent at very short times into the trajectory, if the signal matrix decays very rapidly. The second term contributes to misalignment between the updates as it is always negative, which can be shown via the triangle inequality considering vectors $\vec{v}$ and $\vec{w}=\delta t\vec{z}-t^*\vec{y}$. We validate this approximation through the misalignment $\mathcal{A}\left(\Delta \vec  w^L,\Delta \vec w^G\right)-1$ in Fig.~\ref{fig:CorrelationB}a, which compares the analytical result from Eq.~\eqref{eq:B8} against a numerical simulation of LIF dynamics (Eq.~\eqref{eq:LIF}) in a network with $N=4$ nodes. Each learning degree of freedom was independently initialized from a normal distribution $\mathcal{N}(0,1)$; likewise, the initial activations and target output values were independently sampled component-wise from $ \mathcal{N}(0,1)$. The task was chosen to be allostery-like, imposing the state vector to match the randomized target at time $t^*=0.05$. The approximation closely tracks the numerical results for values up to $\delta t \approx 0.1$ for a fixed time of error $t^*=0.05$. 

We also consider networks of linear oscillators, $\vec{F} = A\vec{x}$. If we further assume a particular dependence of the dynamical matrix on the tunable DOF $A_{ab} = w_{ab}$, we obtain a simple expression for the alignment. Denoting $\vec{u}_c$ as the normalized gradient of the cost $\vec{u}_c=\vec{\nabla} c / \|\vec{\nabla} c\|$ and $\vec{u}_x$ as the normalized state vector at the time of the error $\vec{u}_x=x(t^*)/\|x(t^*)\|$, we decompose $A$ into symmetric ($A_s$) and anti-symmetric ($A_a$) parts and rewrite:
\begin{equation}
\begin{aligned}
&\mathcal{A}(\Delta \vec w^L, \Delta \vec w^G) = 1+ \\&-\frac{1}{8}\left(\delta t-t^*\right)^2\left[\vec{u}_c^TA_s^2\vec{u}_c-\left(\vec{u}_c^TA_s\vec{u}_c\right)^2\right]\\&-\frac{1}{8}\left(\delta t+t^*\right)^2\left[\vec{u}_c^T\left(-A_a^2\right)\vec{u}_c
\right]\\&-\frac{1}{8}\left(\delta t+t^*\right)^2\left[u_x^TA^2u_x-(u_x^T A u_x)^2\right]
\end{aligned}
\label{eq:B9}
\end{equation}

In this linear case, misalignment is driven by three distinct contributions, each governed by a specific temporal scaling. The first two terms arise from the symmetric ($A_s$) and antisymmetric ($A_a$) components of the dynamical matrix, both representing variances along the direction defined by the error vector $\vec{\nabla} c(t^*)$. The symmetric contribution, however, possesses a unique property: it vanishes when the forward and backwards timescales coincide ($\delta t = t^*$). Since the backward propagator is the transpose of the forward one, this term effectively accounts for the mismatch in the temporal reach of the trajectories; physically, this implies that we can eliminate this specific source of error by tuning the learning rule's integration window to match the backwards trajectory timescale. Furthermore, this term vanishes if the variance inequality is saturated; for instance, when $\vec{\nabla} c(t^*)$ is an eigenvector of $A_s$. In contrast, the antisymmetric contribution scales with the sum of the timescales $(\delta t + t^*)^2$ and cannot be neutralized by choosing a particular $\delta t$. Because this contribution to the misalignment depends on the total duration of the forward and backward fluxes rather than their difference, this component remains finite in all non-trivial cases. Given that $A_a^2$ is negative semi-definite, being the square of a skew-symmetric matrix, this contribution is always negative, causing unavoidable discrepancy between the cost gradient and its local approximation except for trivial cases in which the system is perfectly reciprocal. Non-reciprocal (active) dynamics, which are indeed common in biological learning (e.g. synaptic connections in the brain) are thus expected to impede the local approximation of a gradient, yet clear evidence from nature shows that such interactions do not preclude the ability to learn. We leave the further exploration of the role, advantages and limitations of non-reciprocal dynamics in physical learning to an upcoming study. Finally, the third term captures the variance of the full dynamical matrix $A$ in the direction of the system state at the time of the error $\vec{x}(t^*)$. Like the antisymmetric part, this contribution is proportional to the square sum of the timescales and cannot be eliminated by simply adjusting $\delta t$. Interestingly, this term vanishes as the system approaches a fixed point ($\vec{F} = A\vec{x} \approx 0$), suggesting that the heterogeneity of activity across different modes only obstructs learning during transient dynamics, while the local approximation becomes more accurate as the system's dynamics stabilize.

 Fig.~\ref{fig:CorrelationB}b studies the misalignment in the previously discussed linear case for the particular choice $\delta t=t^*$, which cancels the first term in Eq.~\eqref{eq:B9}, for different values of these variables with respect to the percentage of symmetry of the dynamical matrix $A$. The initial condition for the dynamics $x_i(0)$, the target, and the components of the dynamical matrix $A$ were sampled from $\mathcal{N}(0,1)$. The dynamical matrix $A$ was then split in symmetric and antisymmetric components and linearly interpolated following the desired percentage of symmetry $\gamma$; $A_{\gamma}=\left(1-\gamma\right)A_s+\gamma A_a$, in order to avoid issues related to the size of the matrix, all the previous matrices were normalized using their Frobenius norm. As predicted by Eq.~\eqref{eq:B9}, if the symmetric part dominates $A$, the alignment tends to increase, with its maximum determined by the transient dynamics contribution, which almost coincides with the full symmetry in this case. However, when anti-symmetric contributions are important, we see larger misalignment, increasing with the length of the temporal trajectory $\delta t+t^*$.

\section{Synchronization task on a reciprocal Kuramoto network}
\label{sec:recipKuramoto}

\begin{figure*}[t]
\includegraphics[width=0.85\linewidth]{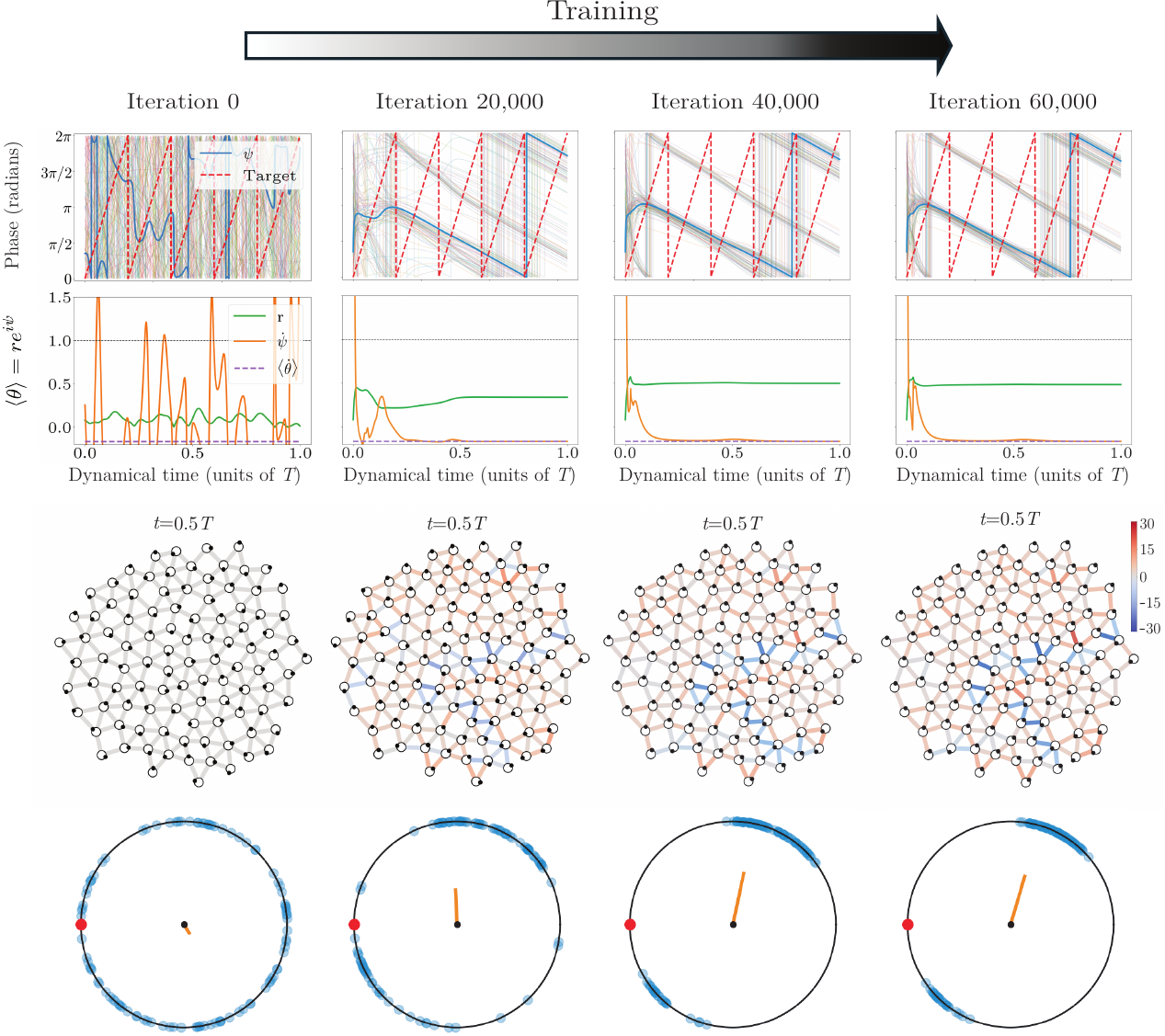}
\caption{Training snapshots for a \textit{reciprocal} network of $N=100$ Kuramoto oscillators spatially organized in 2D, and coupled locally with an average of $Z=4.62$  neighbors (c.f. Fig.~\ref{fig:figKuramoto} of the main text). The oscillators have intrinsic frequencies $\omega_a$ drawn from a normal distribution with mean zero and unit variance and are initially distributed uniformly in phase with vanishing couplings.  Early in training (left column) no synchronization occurs because the oscillators are decoupled. As training progresses (second, third and fourth columns), the couplings adjust and the system \textit{does} learn to congregrate and synchronize, but only at the initialized mean frequency $\langle \dot{\theta} \rangle = \langle\omega\rangle \neq 1$.
({\bf Top row}) Desired phase (red dashed), actual phase of the Kuramoto order parameter $\psi(t)$ (blue), and phases of individual oscillators (faded colors), over four epochs of training.  At the end of training the oscillators largely synchronize \textit{away} from the target frequency at $\langle\omega\rangle \neq 1$.
({\bf Second row})  Magnitude ($r$) and frequency ($\dot{\psi}$) of the Kuramoto order parameter over time in each training epoch.  The magnitude grows with training. but  the frequency does not approach the target value, $\dot{\psi} = \langle \dot{\theta} \rangle = \langle\omega \rangle \neq 1$, indicating synchrony but at the initialized mean intrinsic frequency, not the target frequency. 
({\bf Third row}) Local coupling strengths $K_{ab} = K_{ba}$ (blue = negative, red = positive) between oscillators (circles), indicated at a time $t=T/2$ within each training epoch. The couplings are reciprocal with some growing over time (phases at $t=T/2$ indicated as a black dot in the circle representing each oscillator). 
({\bf Fourth row}) Phases of all oscillators (blue dot), target phase (red dot),
and Kuramoto order parameter $re^{i\psi}$ (orange line, length = r, angle = $\psi$) at a time $t=T/2$ within each training iteration, showing off-target synchronization after training.
}
\label{fig:figReciprocalKuramoto}
\end{figure*}

\begin{figure*}[t]
\includegraphics[width=\linewidth]{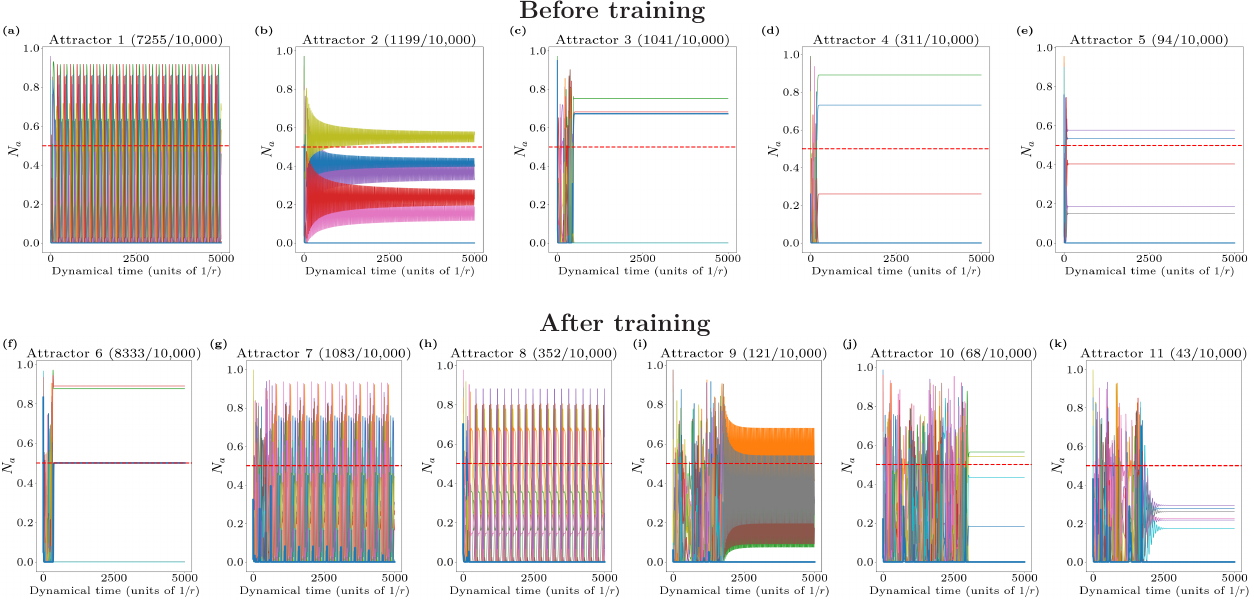}
\caption{The catalogue of attractor states for a system following ecological dynamics, Eq.~\eqref{eq:S_LV_dyn}. We choose 10,000 random initializations for a system of $S=50$ interacting species and then simulate the dynamics for each initialization until the system reaches a dynamical attractor. We do this for the same set of random initializations both before \emph{and} after training. (\textbf{Top row}) Before training, we observe five distinct attractors, plotted in decreasing order of observed frequency. (\textbf{a}) The most frequently observed attractor is a quasiperiodic limit cycle occurring 7255/10,000 times. We note, before training, the initialization repeatedly used during training, $N_{a}(0) = 0.5$, settles into this attractor if simulated long enough. The remaining observed attractors are (\textbf{b}) a periodic limit cycle attractor with frequency 1199/10,000, (\textbf{c}) a fixed point attractor with frequency 1041/10,000, (\textbf{d}) a second distinct fixed point attractor with frequency 311/10,000, and (\textbf{e}) a third distinct fixed point attractor with frequency 94/10,000. The relevant output species in shown in dark blue in all panels; it is only appreciably alive in Attractor 3.
(\textbf{Bottom row}) After training, we observe six distinct attractors, plotted in decreasing order of observed frequency. (\textbf{f}) The most frequently observed attractor is a fixed point attractor occurring 8333/10,000 times. In this fixed point, the target species is exactly on target. The remaining observed attractors are (\textbf{g}) a quasiperiodic limit cycle attractor with frequency 1083/10,000, (\textbf{h}) a distinct quasiperiodic limit cycle attractor with frequency 352/10,000, (\textbf{i}) a periodic limit cycle attractor with frequency 121/10,000, (\textbf{j}) a second distinct fixed point attractor with frequency 68/10,000, and (\textbf{j}) a third distinct fixed point attractor with frequency 43/10,000}
\label{fig:figLV_attractors}
\end{figure*}

In Sec.~\ref{sec:Kuramoto} of the main text, we trained a network of locally coupled Kuramoto oscillators to synchronize at a  frequency away from the average of the intrinsic frequencies of the individual oscillators. We studied a non-reciprocal network and argued this task cannot be accomplished in a reciprocal network. To illustrate the latter argument, here we apply the same training paradigm for a similar network (identical connectivity, initialization, and training hyperparameters) but only train  reciprocal interactions, i.e. we require that $K_{ab} = K_{ba}$ in Eq.~\eqref{eq:Kuramoto_dyn}. We plot the results of training in Fig.~\ref{fig:figReciprocalKuramoto} (compare to Fig.~\ref{fig:figKuramoto}). As stated in the main text, synchronization will not occur in this network because of the structure of the dynamics. We reproduce the argument here. Consider the dynamics (reproduced from Eq.~\eqref{eq:Kuramoto_dyn}:
\begin{equation}
    \dot{\theta}_a = \omega_a + \sum_{b\neq a} K_{ab}\sin(\theta_b-\theta_a)
\end{equation}
Let us now consider the mean phase velocity,
\begin{equation}
    \langle \dot{\theta}_a \rangle = \langle \omega_a\rangle + \bigg\langle  \sum_{b\neq a} K_{ab}\sin(\theta_b-\theta_a)\bigg\rangle
\end{equation}
If we are restricting to the reciprocal case, $K_{ab} = K_{ba}$, such that
\begin{equation}
    \bigg\langle  \sum_{b\neq a} K_{ab}\sin(\theta_b-\theta_a)\bigg\rangle =\frac{1}{N} \sum_{a,b} K_{ab} \sin(\theta_b-\theta_a) = 0
\end{equation}
as terms cancel pairwise: $K_{ab}\sin(\theta_b-\theta_a) = -K_{ba}\sin(\theta_a-\theta_b)$. Therefore, regardless of how we train, we know that $\langle \dot{\theta}\rangle  = \langle \omega \rangle$. We select the intrinsic frequencies $\omega_a \sim \mathcal{N} (0,1)$ such that the mean is away from $\omega_{\text{sync}} = 1$, the synchronization target. 
For  our specific random initialization, $\langle \omega\rangle \approx -0.165$)), and we \textit{do not} adjust the intrinsic frequencies $\omega_a$. Indeed, in Fig.~\ref{fig:figReciprocalKuramoto}, the $\langle\dot{\theta}\rangle$ is constant  over training (dashed purple line in second row). After training, the network does learn to congregate, but given that $\langle\dot{\theta}\rangle$ cannot change, the network synchronizes only at that value $\dot{\psi} \approx \langle\dot{\theta}\rangle$.

No such pairwise cancellation occurs for a non-reciprocal network. Looking back at Fig.~\ref{fig:figKuramoto} of the main text, we see that, although we do not train the intrinsic frequencies and thus $\langle \omega\rangle \approx -0.165$ is fixed, $\langle{\dot{\theta}}\rangle$ \textit{does} change over training. thus allowing the system to synchronize at the target frequency $\omega_{\text{sync}}=1.$

\section{Attractor landscape for ecological dynamics}
\label{app:eco}

In Sec.~\ref{sec:LV} of the main text, we trained a network of interacting species following ecological dynamics to reach a fixed point with a desired target value. In particular, we focused on the dynamics (reproduced from Eq.~\eqref{eq:LV_dynamics}),
\begin{equation}
  \dot{N}_a =  \frac{r_a}{K_a} N_a \left(K_a - N_a - \sum_{b \neq a} ^SA_{ab} N_b \right) + D_a
  \label{eq:S_LV_dyn}
\end{equation}
with the parameter choices $K_a=r_a = 1$ and $D_a = 10^{-6}$, and the interaction matrix initially drawn from the uniform distribution $A_{ab} \sim U(0,2)$. For these parameter choices, the system displays multiple attractor states, which we catalog here. In particular, we sample 10,000 random initial conditions, chosen from $N_a \sim U(0,1)$. For a given set of initial conditions, we allow dynamics to run for $10,000$ time units and classify if the system has reached a fixed point. Fixed points are most reliably classified so, for any initialization that has not yet reached a fixed point, we then we allow the system to run for an additional $10,000$ time units and again verify if the system has reached a fixed point. We repeat this process a total of three times such that any non-fixed-point attractors are run for $T=30,000$ time units. After this process, we observe that all initializations settle into an attractor state and all observed attractors may be classified either as fixed points, periodic limit cycles, or quasiperiodic limit cycles. Before training, we observe five attractors: three distinct fixed points, one periodic limit cycle, and one quasiperiodic limit cycle. We plot them in decreasing order of abundance in the top two row of Fig.~\ref{fig:figLV_attractors}. As we can see, the most frequently encountered attractor state is a quasiperiodic attractor (Fig.~\ref{fig:figLV_attractors}a), which appears in 7255 initializations out of the 10,000 random initializations. Note that the output species has an appreciable abundance (dark blue) in only one of the attractor states, a fixed point with frequency 1041/10,000. In this attractor, the output species has a fixed point value of $N_{\text{out}} = 0.67$.

After training, we observe six attractors: three distinct fixed points, one periodic limit cycle, and two distinct quasiperiodic limit cycles. We plot them in the bottom row of Fig.~\ref{fig:figLV_attractors}. Now, the most frequently encountered attractor state is a fixed point attractor, which appears in 8333 initializations out of the 10,000 random initializations (Fig.~\ref{fig:figLV_attractors}f). In this most frequently encountered attractor, the output species settles exactly on target value $N_{\text{out}} = 0.5$. Using this random sampling, we observe that training effectively reshaped the attractor landscape: the system still exhibits multiple attractors after training (in fact, it exhibits more attractors), but one learned attractor (Attractor 6, Fig.~\ref{fig:figLV_attractors}f) corresponds to the desired behavior and its basin of attraction has enveloped a significant share of the sampled space.

\bibliographystyle{unsrt}
\bibliography{Citations}

@article{mayer2015well,
  title={How a well-adapted immune system is organized},
  author={Mayer, Andreas and Balasubramanian, Vijay and Mora, Thierry and Walczak, Aleksandra M},
  journal={Proceedings of the National Academy of Sciences},
  volume={112},
  number={19},
  pages={5950--5955},
  year={2015},
  publisher={National Academy of Sciences}
}

@article{mayer2016diversity,
  title={Diversity of immune strategies explained by adaptation to pathogen statistics},
  author={Mayer, Andreas and Mora, Thierry and Rivoire, Olivier and Walczak, Aleksandra M},
  journal={Proceedings of the National Academy of Sciences},
  volume={113},
  number={31},
  pages={8630--8635},
  year={2016},
  publisher={National Academy of Sciences}
}

@article{bradde2020size,
  title={The size of the immune repertoire of bacteria},
  author={Bradde, Serena and Nourmohammad, Armita and Goyal, Sidhartha and Balasubramanian, Vijay},
  journal={Proceedings of the National Academy of Sciences},
  volume={117},
  number={10},
  pages={5144--5151},
  year={2020},
  publisher={National Academy of Sciences}
}

@article{hermundstad2014variance,
  title={Variance predicts salience in central sensory processing},
  author={Hermundstad, Ann M and Briguglio, John J and Conte, Mary M and Victor, Jonathan D and Balasubramanian, Vijay and Tka{\v{c}}ik, Ga{\v{s}}per},
  journal={Elife},
  volume={3},
  pages={e03722},
  year={2014},
  publisher={eLife Sciences Publications, Ltd}
}

@article{ratliff2010retina,
  title={Retina is structured to process an excess of darkness in natural scenes},
  author={Ratliff, Charles P and Borghuis, Bart G and Kao, Yen-Hong and Sterling, Peter and Balasubramanian, Vijay},
  journal={Proceedings of the National Academy of Sciences},
  volume={107},
  number={40},
  pages={17368--17373},
  year={2010},
  publisher={National Acad Sciences}
}

@article{garrigan2010design,
  title={Design of a trichromatic cone array},
  author={Garrigan, Patrick and Ratliff, Charles P and Klein, Jennifer M and Sterling, Peter and Brainard, David H and Balasubramanian, Vijay},
  journal={PLoS computational biology},
  volume={6},
  number={2},
  pages={e1000677},
  year={2010},
  publisher={Public Library of Science San Francisco, USA}
}

@article{smith2006efficient,
  title={Efficient auditory coding},
  author={Smith, Evan C and Lewicki, Michael S},
  journal={Nature},
  volume={439},
  number={7079},
  pages={978--982},
  year={2006},
  publisher={Nature Publishing Group UK London}
}

@article{lewicki2002efficient,
  title={Efficient coding of natural sounds},
  author={Lewicki, Michael S},
  journal={Nature neuroscience},
  volume={5},
  number={4},
  pages={356--363},
  year={2002},
  publisher={Nature Publishing Group US New York}
}

@article{tecsileanu2017rules,
  title={Rules and mechanisms for efficient two-stage learning in neural circuits},
  author={Te{\c{s}}ileanu, Tiberiu and {\"O}lveczky, Bence and Balasubramanian, Vijay},
  journal={Elife},
  volume={6},
  pages={e20944},
  year={2017},
  publisher={eLife Sciences Publications, Ltd}
}

@article{tecsileanu2019adaptation,
  title={Adaptation of olfactory receptor abundances for efficient coding},
  author={Te{\c{s}}ileanu, Tiberiu and Cocco, Simona and Monasson, Remi and Balasubramanian, Vijay},
  journal={Elife},
  volume={8},
  pages={e39279},
  year={2019},
  publisher={eLife Sciences Publications, Ltd}
}

@article{krishnamurthy2022disorder,
  title={Disorder and the neural representation of complex odors},
  author={Krishnamurthy, Kamesh and Hermundstad, Ann M and Mora, Thierry and Walczak, Aleksandra M and Balasubramanian, Vijay},
  journal={Frontiers in Computational Neuroscience},
  volume={16},
  pages={917786},
  year={2022},
  publisher={Frontiers}
}

@article{wei2015principle,
  title={A principle of economy predicts the functional architecture of grid cells},
  author={Wei, Xue-Xin and Prentice, Jason and Balasubramanian, Vijay},
  journal={Elife},
  volume={4},
  pages={e08362},
  year={2015},
  publisher={eLife Sciences Publications, Ltd}
}

@book{sterling2015principles,
  title={Principles of neural design},
  author={Sterling, Peter and Laughlin, Simon},
  year={2015},
  publisher={MIT press}
}

@article{shin2017liquid,
  title={Liquid phase condensation in cell physiology and disease},
  author={Shin, Yongdae and Brangwynne, Clifford P},
  journal={Science},
  volume={357},
  number={6357},
  pages={eaaf4382},
  year={2017},
  publisher={American Association for the Advancement of Science}
}

@article{trifonova2025trainable,
  title={Trainable computation in molecular networks},
  author={Trifonova, Kristina and Falk, Martin J and Rouches, Mason and Vaikuntanathan, Suriyanarayanan and Elowitz, Michael and Murugan, Arvind},
  journal={bioRxiv},
  pages={2025--12},
  year={2025},
  publisher={Cold Spring Harbor Laboratory}
}

@article{christensen20222022,
  title={2022 roadmap on neuromorphic computing and engineering},
  author={Christensen, Dennis V and Dittmann, Regina and Linares-Barranco, Bernabe and Sebastian, Abu and Le Gallo, Manuel and Redaelli, Andrea and Slesazeck, Stefan and Mikolajick, Thomas and Spiga, Sabina and Menzel, Stephan and others},
  journal={Neuromorphic Computing and Engineering},
  volume={2},
  number={2},
  pages={022501},
  year={2022},
  publisher={IOP Publishing}
}

@book{dayan2005theoretical,
  title={Theoretical neuroscience: computational and mathematical modeling of neural systems},
  author={Dayan, Peter and Abbott, Laurence F},
  year={2005},
  publisher={MIT press}
}

@article{mehta2019high,
  title={A high-bias, low-variance introduction to machine learning for physicists},
  author={Mehta, Pankaj and Bukov, Marin and Wang, Ching-Hao and Day, Alexandre GR and Richardson, Clint and Fisher, Charles K and Schwab, David J},
  journal={Physics reports},
  year={2019},
  publisher={Elsevier}
}

@article{wang2025mechanosensitive,
  title={Mechanosensitive Remodeling Sustains Rigidity Homeostasis in Actin Cortex Models},
  author={Wang, Haina and Cunha, Marco A Galvani and Crocker, John C and Liu, Andrea J},
  journal={bioRxiv},
  pages={2025--10},
  year={2025},
  publisher={Cold Spring Harbor Laboratory}
}

@inproceedings{bengio2015stdp,
  title={From stdp towards biologically plausible deep learning},
  author={Bengio, Yoshua and Fischer, Asja and Mesnard, Thomas and Zhang, Saizheng and Wu, Yuhai},
  booktitle={Deep Learning Workshop, International Conference on Machine Learning (ICML)},
  year={2015}
}

@article{richards2019deep,
  title={A deep learning framework for neuroscience},
  author={Richards, Blake A and Lillicrap, Timothy P and Beaudoin, Philippe and Bengio, Yoshua and Bogacz, Rafal and Christensen, Amelia and Clopath, Claudia and Costa, Rui Ponte and de Berker, Archy and Ganguli, Surya and others},
  journal={Nature neuroscience},
  volume={22},
  number={11},
  pages={1761--1770},
  year={2019},
  publisher={Nature Publishing Group}
}

@article {stern2020supervised,
	author = {Stern, Menachem and Arinze, Chukwunonso and Perez, Leron and Palmer, Stephanie E. and Murugan, Arvind},
	title = {Supervised learning through physical changes in a mechanical system},
	volume = {117},
	number = {26},
	pages = {14843--14850},
	year = {2020},
	doi = {10.1073/pnas.2000807117},
	publisher = {National Academy of Sciences},
	issn = {0027-8424},
	URL = {https://www.pnas.org/content/117/26/14843},
	eprint = {https://www.pnas.org/content/117/26/14843.full.pdf},
	journal = {Proceedings of the National Academy of Sciences}
}

@article{stern2021supervised,
  title={Supervised learning in physical networks: From machine learning to learning machines},
  author={Stern, Menachem and Hexner, Daniel and Rocks, Jason W and Liu, Andrea J},
  journal={Physical Review X},
  volume={11},
  number={2},
  pages={021045},
  year={2021},
  publisher={APS}
}

@incollection{movellan1991contrastive,
  title={Contrastive Hebbian learning in the continuous Hopfield model},
  author={Movellan, Javier R},
  booktitle={Connectionist models},
  pages={10--17},
  year={1991},
  publisher={Elsevier}
}

@article{scellier2017equilibrium,
  title={Equilibrium propagation: Bridging the gap between energy-based models and backpropagation},
  author={Scellier, Benjamin and Bengio, Yoshua},
  journal={Frontiers in computational neuroscience},
  volume={11},
  pages={24},
  year={2017},
  publisher={Frontiers}
}

@article{lopez2023self,
  title={Self-learning machines based on Hamiltonian echo backpropagation},
  author={Lopez-Pastor, Victor and Marquardt, Florian},
  journal={Physical Review X},
  volume={13},
  number={3},
  pages={031020},
  year={2023},
  publisher={APS}
}

@article{kendall2020training,
  title={Training End-to-End Analog Neural Networks with Equilibrium Propagation},
  author={Kendall, Jack and Pantone, Ross and Manickavasagam, Kalpana and Bengio, Yoshua and Scellier, Benjamin},
  journal={arXiv preprint arXiv:2006.01981},
  year={2020}
}

@article{dillavou2022demonstration,
  title={Demonstration of Decentralized Physics-Driven Learning},
  author={Dillavou, Sam and Stern, Menachem and Liu, Andrea J and Durian, Douglas J},
  journal={Physical Review Applied},
  volume={18},
  number={1},
  pages={014040},
  year={2022},
  publisher={APS}
}

@article{rocks2017designing,
  title={Designing allostery-inspired response in mechanical networks},
  author={Rocks, Jason W and Pashine, Nidhi and Bischofberger, Irmgard and Goodrich, Carl P and Liu, Andrea J and Nagel, Sidney R},
  journal={Proceedings of the National Academy of Sciences},
  volume={114},
  number={10},
  pages={2520--2525},
  year={2017},
  publisher={National Acad Sciences}
}

@article{scurria2026,
      title={Equilibrium Propagation for Non-Conservative Systems}, 
      author={Antonino Emanuele Scurria and Dimitri Vanden Abeele and Bortolo Matteo Mognetti and Serge Massar},
      year={2026},
      eprint={2602.03670},
      journal={arXiv preprint arXiv:2602.03670},
      archivePrefix={arXiv},
      primaryClass={cs.LG},
      url={https://arxiv.org/abs/2602.03670}, 
}

@article{ezraty2025harnessing,
  title = {Harnessing intuitive local evolution rules for physical learning},
  author = {Ezraty, Roie and Stern, Menachem and Rubinstein, Shmuel M},
  journal = {Phys. Rev. E},
  volume = {113},
  issue = {2},
  pages = {025304},
  numpages = {11},
  year = {2026},
  month = {Feb},
  publisher = {American Physical Society},
  doi = {10.1103/51dl-czj3},
  url = {https://link.aps.org/doi/10.1103/51dl-czj3}
}

@article{valiant1984theory,
  title={A theory of the learnable},
  author={Valiant, Leslie G},
  journal={Communications of the ACM},
  volume={27},
  number={11},
  pages={1134--1142},
  year={1984},
  publisher={ACM New York, NY, USA}
}

@article{arzash2025rigidity,
  title={Rigidity of epithelial tissues as a double optimization problem},
  author={Arzash, Sadjad and Tah, Indrajit and Liu, Andrea J and Manning, M Lisa},
  journal={Physical Review Research},
  volume={7},
  number={1},
  pages={013157},
  year={2025},
  publisher={APS}
}

@article{dillavou2025understanding,
  title={Understanding and Embracing Imperfection in Physical Learning Networks},
  author={Dillavou, Sam and Guzman, Marcelo and Liu, Andrea J and Durian, Douglas J},
  journal={arXiv preprint arXiv:2505.22887},
  year={2025}
}

@article{wycoff2022learning,
author = {Wycoff, Jacob F. and Dillavou, Sam  and Stern, Menachem  and Liu, Andrea J.  and Durian, Douglas J.},
title = {Desynchronous learning in a physics-driven learning network},
journal = {The Journal of Chemical Physics},
volume = {156},
number = {14},
pages = {144903},
year = {2022},
doi = {10.1063/5.0084631},
URL = {https://doi.org/10.1063/5.0084631},
eprint = {https://doi.org/10.1063/5.0084631}
}

@article{scellier2018generalization,
  title={Generalization of equilibrium propagation to vector field dynamics},
  author={Scellier, Benjamin and Goyal, Anirudh and Binas, Jonathan and Mesnard, Thomas and Bengio, Yoshua},
  journal={arXiv preprint arXiv:1808.04873},
  year={2018}
}

@article{teeter2018generalized,
  title={Generalized leaky integrate-and-fire models classify multiple neuron types},
  author={Teeter, Corinne and Iyer, Ramakrishnan and Menon, Vilas and Gouwens, Nathan and Feng, David and Berg, Jim and Szafer, Aaron and Cain, Nicholas and Zeng, Hongkui and Hawrylycz, Michael and others},
  journal={Nature communications},
  volume={9},
  number={1},
  pages={709},
  year={2018},
  publisher={Nature Publishing Group UK London}
}

@article{rozenberg2019ultra,
  title={An ultra-compact leaky-integrate-and-fire model for building spiking neural networks},
  author={Rozenberg, MJ and Schneegans, O and Stoliar, P},
  journal={Scientific reports},
  volume={9},
  number={1},
  pages={11123},
  year={2019},
  publisher={Nature Publishing Group UK London}
}

@article{stern2022physical,
  title={Physical learning beyond the quasistatic limit},
  author={Stern, Menachem and Dillavou, Sam and Miskin, Marc Z and Durian, Douglas J and Liu, Andrea J},
  journal={Physical Review Research},
  volume={4},
  number={2},
  pages={L022037},
  year={2022},
  publisher={APS}
}

@article{arinze2023learning,
  title = {Learning to self-fold at a bifurcation},
  author = {Arinze, Chukwunonso and Stern, Menachem and Nagel, Sidney R. and Murugan, Arvind},
  journal = {Phys. Rev. E},
  volume = {107},
  issue = {2},
  pages = {025001},
  numpages = {11},
  year = {2023},
  month = {Feb},
  publisher = {American Physical Society},
  doi = {10.1103/PhysRevE.107.025001},
  url = {https://link.aps.org/doi/10.1103/PhysRevE.107.025001}
}

@article{stern2023learning,
author = {Stern, Menachem and Murugan, Arvind},
title = {Learning Without Neurons in Physical Systems},
journal = {Annual Review of Condensed Matter Physics},
volume = {14},
number = {1},
pages = {417-441},
year = {2023},
doi = {10.1146/annurev-conmatphys-040821-113439},
URL = {https://doi.org/10.1146/annurev-conmatphys-040821-113439},
eprint = {https://doi.org/10.1146/annurev-conmatphys-040821-113439}
}

@article{falk2023learning,
  title={Learning to learn by using nonequilibrium training protocols for adaptable materials},
  author={Falk, Martin J and Wu, Jiayi and Matthews, Ayanna and Sachdeva, Vedant and Pashine, Nidhi and Gardel, Margaret L and Nagel, Sidney R and Murugan, Arvind},
  journal={Proceedings of the National Academy of Sciences},
  volume={120},
  number={27},
  pages={e2219558120},
  year={2023},
  publisher={National Acad Sciences}
}

@article{anisetti2023learning,
  title={Learning by non-interfering feedback chemical signaling in physical networks},
  author={Anisetti, Vidyesh Rao and Scellier, Benjamin and Schwarz, Jennifer M},
  journal={Physical Review Research},
  volume={5},
  number={2},
  pages={023024},
  year={2023},
  publisher={APS}
}

@article{anisetti2024frequency,
  title={Frequency propagation: Multimechanism learning in nonlinear physical networks},
  author={Anisetti, Vidyesh Rao and Kandala, Ananth and Scellier, Benjamin and Schwarz, JM},
  journal={Neural Computation},
  volume={36},
  number={4},
  pages={596--620},
  year={2024},
  publisher={MIT Press One Rogers Street, Cambridge, MA 02142-1209, USA journals-info~…}
}

@article{dillavou2024machine,
author = {Sam Dillavou  and Benjamin D. Beyer  and Menachem Stern  and Andrea J. Liu  and Marc Z. Miskin  and Douglas J. Durian },
title = {Machine learning without a processor: Emergent learning in a nonlinear analog network},
journal = {Proceedings of the National Academy of Sciences},
volume = {121},
number = {28},
pages = {e2319718121},
year = {2024},
doi = {10.1073/pnas.2319718121},
URL = {https://www.pnas.org/doi/abs/10.1073/pnas.2319718121},
eprint = {https://www.pnas.org/doi/pdf/10.1073/pnas.2319718121},
}

@article{altman2024experimental,
  title={Experimental demonstration of coupled learning in elastic networks},
  author={Altman, Lauren E and Stern, Menachem and Liu, Andrea J and Durian, Douglas J},
  journal={Physical Review Applied},
  volume={22},
  number={2},
  pages={024053},
  year={2024},
  publisher={APS}
}

@article{stern2024physical,
  title={Physical effects of learning},
  author={Stern, Menachem and Liu, Andrea J and Balasubramanian, Vijay},
  journal={Physical Review E},
  volume={109},
  number={2},
  pages={024311},
  year={2024},
  publisher={APS}
}

@article{stern2024training,
  title={Training self-learning circuits for power-efficient solutions},
  author={Stern, Menachem and Dillavou, Sam and Jayaraman, Dinesh and Durian, Douglas J and Liu, Andrea J},
  journal={APL Machine Learning},
  volume={2},
  number={1},
  year={2024},
  publisher={AIP Publishing}
}

@article{Guzman2024Imprints,
  title = {Microscopic Imprints of Learned Solutions in Tunable Networks},
  author = {Guzman, Marcelo and Martins, Felipe and Stern, Menachem and Liu, Andrea J.},
  journal = {Phys. Rev. X},
  volume = {15},
  issue = {3},
  pages = {031056},
  numpages = {19},
  year = {2025},
  month = {Aug},
  publisher = {American Physical Society},
  doi = {10.1103/f2hb-c9s1},
}

@article{Arzash2025Epithelial,
  title = {Epithelial convergent extension as a tuning process},
  url = {http://dx.doi.org/10.1101/2025.11.06.687029},
  DOI = {10.1101/2025.11.06.687029},
  publisher = {openRxiv},
  author = {Arzash,  Sadjad and Liu,  Andrea J. and Manning,  M. Lisa},
  year = {2025},
  month = nov,
  journal = {bioRxiv}
}

@article{audiomnist2023,
    title = {AudioMNIST: Exploring Explainable Artificial Intelligence for audio analysis on a simple benchmark},
    journal = {Journal of the Franklin Institute},
    year = {2023},
    issn = {0016-0032},
    doi = {https://doi.org/10.1016/j.jfranklin.2023.11.038},
    url = {https://www.sciencedirect.com/science/article/pii/S0016003223007536},
    author = {Sören Becker and Johanna Vielhaben and Marcel Ackermann and Klaus-Robert Müller and Sebastian Lapuschkin and Wojciech Samek},
}

@article{hodgkin1952quantitative,
  title={A quantitative description of membrane current and its application to conduction and excitation in nerve},
  author={Hodgkin, Alan L and Huxley, Andrew F},
  journal={The Journal of physiology},
  volume={117},
  number={4},
  pages={500},
  year={1952}
}

@article{acebron2005kuramoto,
  title={The Kuramoto model: A simple paradigm for synchronization phenomena},
  author={Acebr{\'o}n, Juan A and Bonilla, Luis L and P{\'e}rez Vicente, Conrad J and Ritort, F{\'e}lix and Spigler, Renato},
  journal={Reviews of modern physics},
  volume={77},
  number={1},
  pages={137--185},
  year={2005},
  publisher={APS}
}

@article{nahmias2013leaky,
  title={A leaky integrate-and-fire laser neuron for ultrafast cognitive computing},
  author={Nahmias, Mitchell A and Shastri, Bhavin J and Tait, Alexander N and Prucnal, Paul R},
  journal={IEEE journal of selected topics in quantum electronics},
  volume={19},
  number={5},
  pages={1--12},
  year={2013},
  publisher={IEEE}
}

@article{du2025metamaterials,
  title={Metamaterials that learn to change shape},
  author={Du, Yao and Veenstra, Jonas and van Mastrigt, Ryan and Coulais, Corentin},
  journal={arXiv preprint arXiv:2501.11958},
  year={2025}
}

@article{mandal2024learning,
  title={Learning dynamical behaviors in physical systems},
  author={Mandal, Rituparno and Huang, Rosalind and Fruchart, Michel and Moerman, Pepijn G and Vaikuntanathan, Suriyanarayanan and Murugan, Arvind and Vitelli, Vincenzo},
  journal={arXiv preprint arXiv:2406.07856},
  year={2024}
}

@article{massar2025equilibrium,
  title={Equilibrium propagation for learning in Lagrangian dynamical systems},
  author={Massar, Serge},
  journal={Physical Review E},
  volume={112},
  number={3},
  pages={035304},
  year={2025},
  publisher={APS}
}

@article{floyd2024learning,
  title={Learning to control non-equilibrium dynamics using local imperfect gradients},
  author={Floyd, Carlos and Dinner, Aaron R and Vaikuntanathan, Suriyanarayanan},
  journal={arXiv preprint arXiv:2404.03798},
  year={2024}
}

@article{fernandez2024ornstein,
  title={Ornstein--Uhlenbeck Adaptation as a Mechanism for Learning in Brains and Machines},
  author={Fern{\'a}ndez, Jes{\'u}s Garc{\'\i}a and Ahmad, Nasir and van Gerven, Marcel},
  journal={Entropy},
  volume={26},
  number={12},
  pages={1125},
  year={2024}
}

@article{van2026dynamical,
  title={Dynamical systems foundations for neuromorphic intelligence},
  author={van Gerven, Marcel AJ},
  journal={Neuromorphic Computing and Engineering},
  year={2026}
}

@article{banerjee2025learning,
  title={Learning via mechanosensitivity and activity in cytoskeletal networks},
  author={Banerjee, Deb S and Falk, Martin J and Gardel, Margaret L and Walczak, Aleksandra M and Mora, Thierry and Vaikuntanathan, Suriyanarayanan},
  journal={arXiv preprint arXiv:2504.15107},
  year={2025}
}

@article{stern2025physical,
  title={Physical networks become what they learn},
  author={Stern, Menachem and Guzman, Marcelo and Martins, Felipe and Liu, Andrea J and Balasubramanian, Vijay},
  journal={Physical Review Letters},
  volume={134},
  number={14},
  pages={147402},
  year={2025},
  publisher={APS}
}

@article{folse2010individual,
  title={What is an individual organism? A multilevel selection perspective},
  author={Folse III, Henri J and Roughgarden, Joan},
  journal={The Quarterly review of biology},
  volume={85},
  number={4},
  pages={447--472},
  year={2010},
  publisher={The University of Chicago Press}
}

@article{kandel1992biological,
  title={The biological basis of learning and individuality},
  author={Kandel, Eric R and Hawkins, Robert D},
  journal={Scientific American},
  volume={267},
  number={3},
  pages={78--87},
  year={1992},
  publisher={JSTOR}
}

@article{marbach2023vein,
  title={Vein fate determined by flow-based but time-delayed integration of network architecture},
  author={Marbach, Sophie and Ziethen, Noah and Bastin, Leonie and B{\"a}uerle, Felix K and Alim, Karen},
  journal={Elife},
  volume={12},
  pages={e78100},
  year={2023},
  publisher={eLife Sciences Publications Limited}
}

@article{le2023physarum,
  title={Physarum polycephalum: Smart network adaptation},
  author={Le Verge-Serandour, Mathieu and Alim, Karen},
  journal={Annual Review of Condensed Matter Physics},
  volume={15},
  year={2023},
  publisher={Annual Reviews}
}

@article{whiting2016towards,
  title={Towards a Physarum learning chip},
  author={Whiting, James GH and Jones, Jeff and Bull, Larry and Levin, Michael and Adamatzky, Andrew},
  journal={Scientific reports},
  volume={6},
  number={1},
  pages={19948},
  year={2016},
  publisher={Nature Publishing Group UK London}
}

@article{caporale2008spike,
  title={Spike timing--dependent plasticity: a Hebbian learning rule},
  author={Caporale, Natalia and Dan, Yang},
  journal={Annu. Rev. Neurosci.},
  volume={31},
  number={1},
  pages={25--46},
  year={2008},
  publisher={Annual Reviews}
}

@misc{rumelhart1985learning,
  title={Learning internal representations by error propagation},
  author={Rumelhart, David E and Hinton, Geoffrey E and Williams, Ronald J and others},
  year={1985},
  publisher={Institute for Cognitive Science, University of California, San Diego La~…}
}

@inproceedings{carreira2005contrastive,
  title={On contrastive divergence learning},
  author={Carreira-Perpinan, Miguel A and Hinton, Geoffrey},
  booktitle={International workshop on artificial intelligence and statistics},
  pages={33--40},
  year={2005},
  organization={PMLR}
}

@article{li2024training,
  title={Training all-mechanical neural networks for task learning through in situ backpropagation},
  author={Li, Shuaifeng and Mao, Xiaoming},
  journal={Nature Communications},
  volume={15},
  number={1},
  pages={1--12},
  year={2024},
  publisher={Nature Publishing Group}
}

@article{wanjura2024quantum,
  title={Quantum Equilibrium Propagation for efficient training of quantum systems based on Onsager reciprocity},
  author={Wanjura, Clara C and Marquardt, Florian},
  journal={arXiv preprint arXiv:2406.06482},
  year={2024}
}

@article{laydevant2024training,
  title={Training an ising machine with equilibrium propagation},
  author={Laydevant, J{\'e}r{\'e}mie and Markovi{\'c}, Danijela and Grollier, Julie},
  journal={Nature Communications},
  volume={15},
  number={1},
  pages={3671},
  year={2024},
  publisher={Nature Publishing Group UK London}
}

@article{needleman2017active,
  title={Active matter at the interface between materials science and cell biology},
  author={Needleman, Daniel and Dogic, Zvonimir},
  journal={Nature reviews materials},
  volume={2},
  number={9},
  pages={1--14},
  year={2017},
  publisher={Nature Publishing Group}
}

@article{ramaswamy2010mechanics,
  title={The mechanics and statistics of active matter},
  author={Ramaswamy, Sriram},
  journal={Annu. Rev. Condens. Matter Phys.},
  volume={1},
  number={1},
  pages={323--345},
  year={2010},
  publisher={Annual Reviews}
}

@article{shaebani2020computational,
  title={Computational models for active matter},
  author={Shaebani, M Reza and Wysocki, Adam and Winkler, Roland G and Gompper, Gerhard and Rieger, Heiko},
  journal={Nature Reviews Physics},
  volume={2},
  number={4},
  pages={181--199},
  year={2020},
  publisher={Nature Publishing Group UK London}
}

@article{michaelis1913kinetik,
  title={Die kinetik der invertinwirkung},
  author={Michaelis, Leonor and Menten, Maud L and others},
  journal={Biochem. z},
  volume={49},
  number={333-369},
  pages={352},
  year={1913},
  publisher={Berlin}
}

@article{shankar2022topological,
  title={Topological active matter},
  author={Shankar, Suraj and Souslov, Anton and Bowick, Mark J and Marchetti, M Cristina and Vitelli, Vincenzo},
  journal={Nature Reviews Physics},
  volume={4},
  number={6},
  pages={380--398},
  year={2022},
  publisher={Nature Publishing Group UK London}
}

@article{fruchart2021non,
  title={Non-reciprocal phase transitions},
  author={Fruchart, Michel and Hanai, Ryo and Littlewood, Peter B and Vitelli, Vincenzo},
  journal={Nature},
  volume={592},
  number={7854},
  pages={363--369},
  year={2021},
  publisher={Nature Publishing Group UK London}
}

@article{berneman2024designing,
  title={Designing precise dynamical steady states in disordered networks},
  author={Berneman, Marc and Hexner, Daniel},
  journal={arXiv preprint arXiv:2409.05060},
  year={2024}
}

@article{ellenberger2025backpropagation,
  title={Backpropagation through space, time and the brain},
  author={Ellenberger, Benjamin and Haider, Paul and Benitez, Federico and Jordan, Jakob and Max, Kevin and Jaras, Ismael and Kriener, Laura and Petrovici, Mihai A},
  journal={Nature Communications},
  year={2025},
  publisher={Nature Publishing Group UK London}
}

@article{berneman2025equilibrium,
  title={Equilibrium Propagation for Periodic Dynamics},
  author={Berneman, Marc and Hexner, Daniel},
  journal={arXiv preprint arXiv:2506.20402},
  year={2025}
}

@article{altman2025collective,
  title={Collective Hysteron Behavior in Nonlinear Flow Networks},
  author={Altman, Lauren E and Awad, Nadia and Durian, Douglas J and Ruiz-Garcia, Miguel and Katifori, Eleni},
  journal={arXiv preprint arXiv:2502.05570},
  year={2025}
}

@article{chatterjee2025hierarchical,
  title={Hierarchical Loop Stabilization in Periodically Driven Elastic Networks},
  author={Chatterjee, Purba and Katifori, Eleni},
  journal={arXiv preprint arXiv:2503.19681},
  year={2025}
}

@article{pourcel2025learning,
  title={Learning long range dependencies through time reversal symmetry breaking},
  author={Pourcel, Guillaume and Ernoult, Maxence},
  journal={arXiv preprint arXiv:2506.05259},
  year={2025}
}

@article{pourcel2025lagrangian,
  title={Lagrangian-based Equilibrium Propagation: generalisation to arbitrary boundary conditions \& equivalence with Hamiltonian Echo Learning},
  author={Pourcel, Guillaume and Basu, Debabrota and Ernoult, Maxence and Gilra, Aditya},
  journal={arXiv preprint arXiv:2506.06248},
  year={2025}
}

@BOOK{Schwartz2013QFT,
  title     = "Quantum field theory and the standard model",
  author    = "Schwartz, Matthew D",
  publisher = "Cambridge University Press",
  month     =  dec,
  year      =  2013,
  address   = "Cambridge, England",
  language  = "en"
}

@INCOLLECTION{Kuramoto1979Kuramoto,
  title     = "Self-entrainment of a population of coupled non-linear
               oscillators",
  booktitle = "International Symposium on Mathematical Problems in Theoretical
               Physics",
  series = "Lecture Notes in Physics",
  volume = "39",
  author    = "Kuramoto, Yoshiki",
  editor = "H. Araki",
  publisher = "Springer-Verlag",
  year      =  1979
}

@BOOK{Kuramoto1984Chemical,
  title     = "Chemical oscillations, waves, and turbulence",
  author    = "Kuramoto, Y",
  publisher = "Springer",
  series    = "Springer Series in Synergetics",
  edition   =  1984,
  month     =  sep,
  year      =  1984,
  address   = "Berlin, Germany",
  language  = "en"
}

@ARTICLE{Willems1972Reciprocal,
  title     = "Dissipative dynamical systems Part {II}: Linear systems with
               quadratic supply rates",
  author    = "Willems, Jan C",
  journal   = "Arch. Ration. Mech. Anal.",
  publisher = "Springer Science and Business Media LLC",
  volume    =  45,
  number    =  5,
  pages     = "352--393",
  year      =  1972,
  language  = "en"
}

@book{Enderton1972Logic,
	address = {New York,},
	author = {Herbert Bruce Enderton},
	editor = {},
	publisher = {Academic Press},
	title = {A Mathematical Introduction to Logic},
	year = {1972}
}

@Article{Erbas2018MolecularLogic,
author ="Erbas-Cakmak, Sundus and Kolemen, Safacan and Sedgwick, Adam C. and Gunnlaugsson, Thorfinnur and James, Tony D. and Yoon, Juyoung and Akkaya, Engin U.",
title  ="Molecular logic gates: the past{,} present and future",
journal  ="Chem. Soc. Rev.",
year  ="2018",
volume  ="47",
issue  ="7",
pages  ="2228-2248",
publisher  ="The Royal Society of Chemistry",
doi  ="10.1039/C7CS00491E",
}

@article{Gore2022LV,
  title={Emergent phases of ecological diversity and dynamics mapped in microcosms},
  author={Hu, Jiliang and Amor, Daniel R. and Barbier, Matthieu and Bunin, Guy and Gore, Jeff},
  journal={Science},
  volume={378},
  number={6615},
  pages={85--89},
  year={2022},
  publisher={American Association for the Advancement of Science},
  doi={10.1126/science.abm7841},
}

@article{Bunin2017LV,
  title = {Ecological communities with Lotka-Volterra dynamics},
  author = {Bunin, Guy},
  journal = {Phys. Rev. E},
  volume = {95},
  issue = {4},
  pages = {042414},
  numpages = {8},
  year = {2017},
  month = {Apr},
  publisher = {American Physical Society},
  doi = {10.1103/PhysRevE.95.042414},
  url = {https://link.aps.org/doi/10.1103/PhysRevE.95.042414}
}

@article{Lotka_Volterra,
author = {Alfred J. Lotka },
title = {Analytical Note on Certain Rhythmic Relations in Organic Systems},
journal = {Proceedings of the National Academy of Sciences},
volume = {6},
number = {7},
pages = {410-415},
year = {1920},
doi = {10.1073/pnas.6.7.410},
URL = {https://www.pnas.org/doi/abs/10.1073/pnas.6.7.410},
eprint = {https://www.pnas.org/doi/pdf/10.1073/pnas.6.7.410}}

\end{document}